\documentclass[twocolumn]{jpsj3}
\usepackage{amssymb}
\usepackage{color}  

\def\lsco{La$_{2-x}$Sr$_x$CuO$_4$}
\def\lbco{La$_{2-x}$Ba$_x$CuO$_4$}

\def\lesco{La$_{1.8-x}$Eu$_{0.2}$Sr$_x$CuO$_4$}
\def\ybco{YBa$_2$Cu$_3$O$_{6+x}$}
\def\nccoc{Ca$_{2-x}$Na$_x$CuO$_2$Cl$_2$}
\def\lbcoate{La$_{1.875}$Ba$_{0.125}$CuO$_{4}$}
\def\bscco{Bi$_2$Sr$_2$CaCu$_2$O$_{8+\delta}$}
\def\plcco{Pr$_{1-x}$LaCe$_x$CuO$_{4+\delta}$}
\def\ncco{Nd$_{2-x}$Ce$_x$CuO$_4$}

\def\qaf{${\bf Q}_{\rm AF}$}
\def\ecross{$E_{\rm cross}$}

\title{Progress in Neutron Scattering Studies of \\ Spin Excitations in High-$T_c$ Cuprates }

\author{Masaki Fujita$^1$,  Haruhiro Hiraka$^1$, Masaaki Matsuda,$^2$ Masato Matsuura$^1$, John M. Tranquada,$^3$ Shuichi Wakimoto,$^4$ Guangyong Xu,$^3$ and Kazuyoshi Yamada$^{1,5}$
 }
\inst{$^1$Institute for Materials Research, Tohoku University, Katahira, Sendai 980-8577, Japan \\
$^2$Neutron Scattering Science Division, Oak Ridge National Laboratory,\\ Oak Ridge, Tennessee 37831, USA \\
$^3$Condensed Matter Physics \&\ Materials Science Department, Brookhaven National Laboratory, Upton, NY 11973-5000, USA \\
$^4$Quantum Beam Science Directorate, Japan Atomic Energy Agency, Tokai, Ibaraki 319-1195, Japan \\
$^5$WPI Advanced Institute for Materials Research, Tohoku University, Katahira, Aoba-ku, Sendai 980-8577, Japan
}
\abst{Neutron scattering experiments continue to improve our knowledge of spin fluctuations in layered cuprates, excitations that are symptomatic of the electronic correlations underlying high-temperature superconductivity.  Time-of-flight spectrometers, together with new and varied single crystal samples, have provided a more complete characterization of the magnetic energy spectrum and its variation with carrier concentration.  While the spin excitations appear anomalous in comparison with simple model systems, there is clear consistency among a variety of cuprate families.   Focusing initially on hole-doped systems, we review the nature of the magnetic spectrum, and variations in magnetic spectral weight with doping.  We consider connections with the phenomena of charge and spin stripe order, and the potential generality of such correlations as suggested by studies of magnetic-field and impurity induced order.  We contrast the behavior of the hole-doped systems with the trends found in the electron-doped superconductors.  Returning to hole-doped cuprates, studies of translation-symmetry-preserving magnetic order are discussed, along with efforts to explore new systems.  We conclude with a discussion of future challenges. }

\kword{superconductivity, neutron scattering, cuprates}

\begin{document}
\maketitle

\section{Introduction}

It is now 25 years since the remarkable discovery of high-temperature superconductivity by Bednorz and M\"uller,\cite{bedn86} and this year is also the centennial anniversary of the original discovery of superconductivity in the lab of Kammerlingh Onnes.  Possible polaronic effects motivated the initial decision to look at cuprates,\cite{bedn88} but a role for magnetism soon became clear.  Anderson\cite{ande87} pointed out that undoped cuprates should be Mott insulators, with antiferromagnetism driven by superexchange, and antiferromagnetic order in La$_2$CuO$_{4+y}$ was quickly identified by neutron diffraction.\cite{vakn87}  The possibility that an unconventional pairing mechanism involving antiferromagnetic spin fluctuations\cite{scal86,miya86} might be driving high-temperature superconductivity quickly became an important research theme.

Over the last quarter century, a great deal of progress has been made in characterizing the magnetic correlations in cuprate superconductors.  Our focus here is on neutron scattering studies, but there has also been much complementary work with techniques such as nuclear magnetic resonance (NMR) and muon spin rotation ($\mu$SR).  Much of the progress in neutron work is due to the gradual improvement in sample quality, and especially the growth of large, high-quality single crystals.  Another factor has been the development at spallation sources of time-of-flight spectroscopy with two-dimensional, position-sensitive detectors, which complements the use of triple-axis spectrometers at reactor sources.\cite{tran10}

There have already been a number of reviews of neutron scattering work on cuprates,\cite{kast98,bour98,maso01,lynn01,tran07} as well as on relevant theoretical work.\cite{kive03,deml04,lee06,esch06,ogat08}  Two of us were involved in a review published in this same venue five years ago.\cite{birg06}  Here we will emphasize work done since then.

The rest of this article is organized as follows.  The next section provides some relevant background information.  For the class of hole-doped cuprates, we discuss the magnetic excitation spectrum in \S\ref{sc:magex}, we draw connections with the phenomenology of stripe order in  \S\ref{sc:stripes}, and we consider the impact of impurities on the Cu site in \S\ref{sc:impurity}.  Section~\ref{sc:edope} covers electron-doped cuprates, Pr$_{1-x}$LaCe$_x$CuO$_{4+\delta}$ (PLCCO) in particular.  In \S\ref{sc:other}, we review evidence for exotic magnetic order associated with the pseudogap phase, and efforts to search for magnetic excitations in new systems, such as Bi$_{2+x}$Sr$_{2-x}$CuO$_{6+\delta}$.  We conclude in \S\ref{sc:sum} with a summary and consideration of future challenges.

\section{Background}
\label{sc:back}

The common structural feature of all cuprates is the CuO$_2$ plane.  Within these layers, the Cu atoms form an approximately square lattice, with O atoms at the bridging positions between nearest-neighbor Cu's.  It is often sufficient to consider a tetragonal, or pseudo-tetragonal, coordinate system, with $a\sim3.8$~\AA\ corresponding to the nearest-neighbor Cu-Cu spacing.  With such a choice, we can express wave vectors in terms of the reciprocal lattice units (rlu) $(2\pi/a,2\pi/a,2\pi/c)$.  

\begin{figure}[t]
\begin{center}
\includegraphics[width=8.6cm]{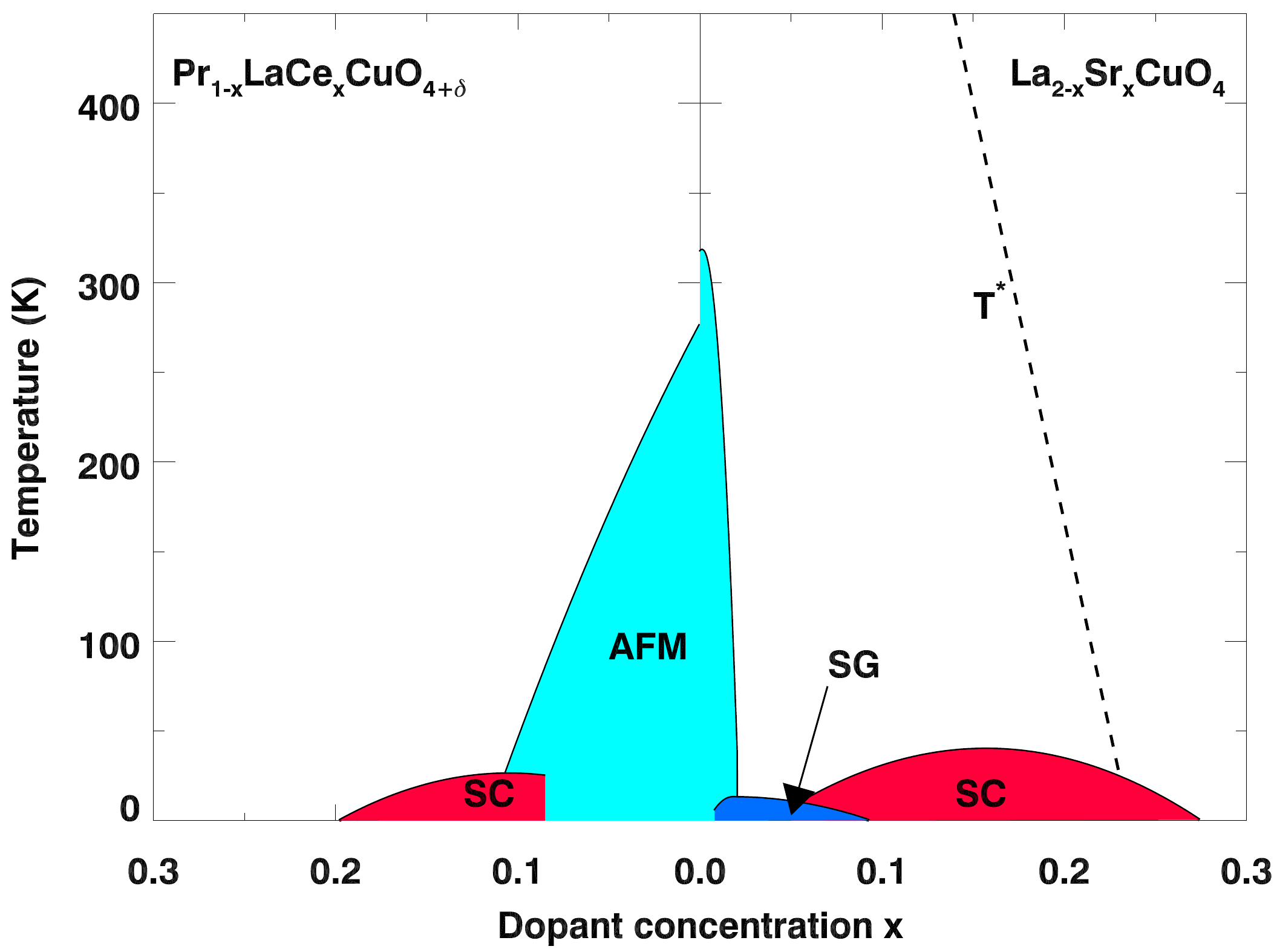} 
 \end{center}
 \caption{(Color online) Representative phase diagrams for hole-doped \lsco\ and electron-doped \plcco, showing antiferromagnetic (AFM), superconducting (SC), and spin-glass (SG) phases.  Below the crossover line labelled $T^\ast$ is the pseudogap regime.}
\label{fg:back}
\end{figure}

Figure~\ref{fg:back} shows representative phase diagrams for hole- and electron-doped cuprates.  In the undoped state, we have an antiferromagnetic insulator.  Each Cu atom has a half-filled $3d_{x^2-y^2}$ orbital, corresponding to a spin $S=1/2$.  The nearest-neighbor spins are coupled by the superexchange interaction, with magnitude $J\sim0.1$~eV.  Antiferromagnetic order doubles the size of the unit cell relative to the tetragonal phase, and, within the planes, is characterized by the ordering wave vector ${\bf Q}_{\rm AF} = (\frac12,\frac12,0)$.  [Theorists often combine this with the rlu, but with $a$ set to 1, so that ${\bf Q}_{\rm AF}=(\pi,\pi)$ in two dimensions.]  Inelastic neutron scattering studies of La$_2$CuO$_4$ show that the magnetic excitations are fairly well described by spin-wave theory,\cite{hayd91,hayd96a,cold01,head10} with the excitations extending over a band width of $\sim2J$.  For \ybco, which has a CuO$_2$ bilayer in each unit cell, one distinguishes between spin excitations that are in-phase between neighboring layers (acoustic mode) or out-of phase (optical mode).

There are some cases where we have to take account of structural deviations from tetragonal.  For \lsco\ (LSCO), the structure over much of the phase diagram is the so-called low-temperature orthorhombic (LTO) phase, in which the unit cell volume is doubled, with a rotation of the in-plane principal axes by 45$^\circ$.  In this structure, the not-quite-orthogonal Cu-O bonds are equivalent to one another, but the antiferromagnetic wave vectors $(1,0,0)_{\rm o}$ and $(0,1,0)_{\rm o}$ are not.  (We will label wave vectors in the LTO coordinate system with a subscript ``o''.)  It is important to distinguish the LTO phase of LSCO from the orthorhombic structure of superconducting \ybco\ (YBCO).  In the latter, the unit cell size is the same as that of the tetragonal, but Cu-O bonds along the $a$ axis are shorter than those along the $b$ axis.

Looking at Fig.~\ref{fg:back}, one can see that long-range antiferromagnetic (AF) order is destroyed by a very small concentration (2\%) of doped holes.  With further doping, a spin-glass phase develops at low temperatures and coexists with the superconducting (SC) phase for $x\gtrsim0.06$ in LSCO.\cite{kast98,birg06}  The doping level corresponding to the maximum SC transition temperature, $T_c$, is referred to as ``optimal''.  Below the crossover line indicated as $T^\ast$, one has the so-called pseudogap phase.  With electron-doping the situation is somewhat different.  The AF ordering temperature, $T_{\rm N}$, decreases more slowly with carrier concentration, and the superconducting phase appears with maximum $T_c$ near the point where $T_{\rm N}\rightarrow0$.

In the hole-doped cuprates, the spin excitations are significantly modified by the presence of charge carriers, as we will discuss in the next section.  For samples in the vicinity of optimal doping, there is a redistribution of magnetic spectral weight  on cooling through $T_c$.  A spin gap opens, and weight appears in a ``resonance'' peak above it.\cite{bour98,dai01,chri04,tran04b,pail06}  It has been argued\cite{yu09} that the ratio of the resonance energy to the superconducting gap energy has a universal value of 1.3.

\section{Magnetic excitations}
\label{sc:magex}

\subsection{Universal magnetic spectrum}

\begin{figure}
\begin{center}
\includegraphics[width=8.2cm]{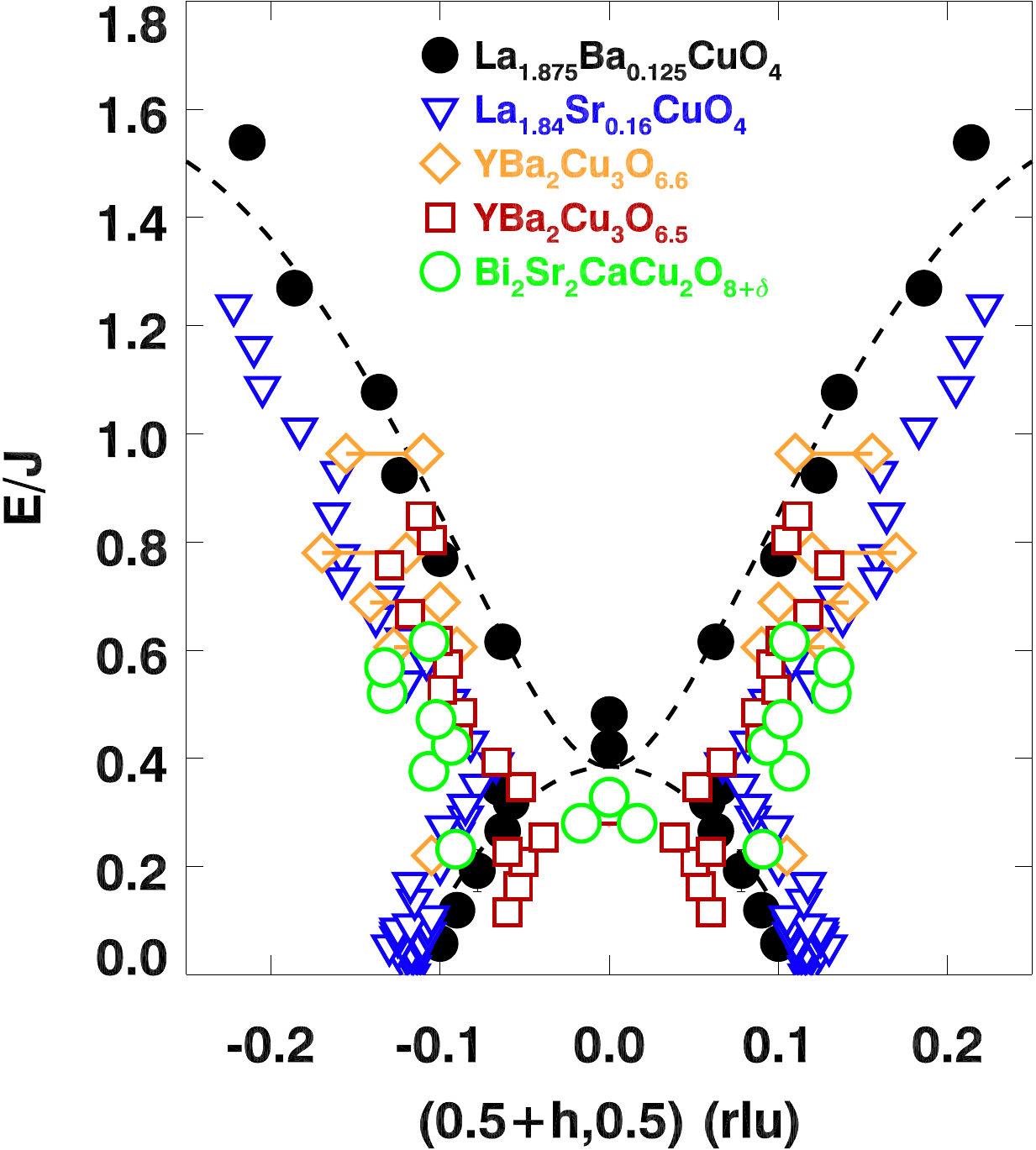}
\end{center}
\caption{(Color online)  Magnetic dispersion relation along $(0.5+h, 0.5, 0)_{\rm tetra}$ in various cuprates, corresponding to wave vectors parallel to the Cu-O bonds.  Data are for \lbcoate,\cite{tran04} La$_{1.84}$Sr$_{0.16}$CuO$_4$,\cite{vign07} YBa$_2$Cu$_3$O$_{6.6}$,\cite{hayd04} YBa$_2$Cu$_3$O$_{6.5}$,\cite{stoc05,stoc10} and \bscco.\cite{xu09} The energy is scaled by $J$ for the AF parent material.\cite{tran07,suga03}}
\label{fg:hg1}
\end{figure}

The inelastic-neutron-scattering (INS) technique using high-flux pulsed neutrons has unveiled the overall spin dynamics of hole-doped cuprates, which extends over a large energy scale, comparable to the bandwidth of $\sim$$2J$ found for the AF parent phase.  While doping causes substantial changes to the magnetic spectrum, a consistent pattern has been identified, as the ``hour-glass'' dispersion was first established in the acoustic magnetic excitations of YBa$_2$Cu$_3$O$_{6.6}$ \cite{hayd04} and in \lbcoate \cite{tran04}.  A comparison of the effective magnetic dispersions about \qaf\ for several different cuprate families is shown in Fig.~\ref{fg:hg1}.  

The spectrum can be thought of in terms of two components, separated by an energy \ecross\ at the waist of the hour glass.  (We note that, while the {\bf q} width of the magnetic scattering is smallest at \ecross, it is a matter of taste whether one describes the {\bf q} dependence there as a single commensurate peak or a set of unresolved incommensurate peaks.) The upwardly dispersing portion, above \ecross, looks similar to what one would expect from AF spin fluctuations with a finite gap, and it is relevant to note that the results for different cuprate families appear to scale with $J$ for the parent AF insulators.  If one considers a fixed excitation energy and looks at the distribution of spectral about \qaf, it is rather isotopic,\cite{stoc05,stoc10,vign07,kofu07} with a possible diamond shape having points oriented along [110] and $[1\bar{1}0]$ directions.\cite{hayd04,tran04,hink07,lips09}  Below \ecross, there appears to be a downward dispersion.  A series of studies on detwinned single crystals of \ybco\ indicate that, at least for underdoped samples, the distribution of spectral weight is quite anisotropic, with the dispersion effectively occurring only along the [100] direction.\cite{hink04,hink07,hink08,haug10,hink10} Such an anisotropy cannot be identified in underdoped LSCO within the superconducting regime, as there is no structural anisotropy between the two Cu-O bond directions; however,  the low-energy excitations disperse down to the incommensurate wave vectors $(\frac{1}{2}\pm\delta, \frac{1}{2},0)$ and $(\frac{1}{2},\frac{1}{2}\pm\delta,0)$ at $E=0$.  We will consider the connection with stripe order in \S\ref{sc:stripes}.

To demonstrate that the hour-glass dispersion is universal, one would like to observe it in cuprate families beyond LSCO and YBCO.  The system \bscco\ (Bi2212) is of particular interest, as it is the prototypical system for studies with angle-resolved photoemission spectroscopy (ARPES) and scanning tunneling spectroscopy (STS).  The challenge has been to grow crystals of sufficient size for inelastic neutron scattering studies.  Initial experiments on smaller crystals allowed one to observe the temperature-dependent development of the resonance peak below $T_c$,\cite{fong99,he01,capo07,mats09c} but identifying the dispersion was a challenge.\cite{fauq07}
Recently, large crystals of optimally-doped were successfully grown,\cite{wen08} enabling the direct measurement of the magnetic excitations.\cite{xu09}  The effective dispersion is indicated by the open circles in Fig.~\ref{fg:hg1}.  The good consistency with the other systems suggests that this behavior is universal.


\subsection{Impact of doped holes}

While the excitations dispersing upwardly from \ecross\ appear to evolve into the AF spin waves as doping is reduced, the downwardly-dispersing excitations clearly reflect the impact of the doped holes, as they have no simple correspondence with any features of the parent AF.  The temperature dependence of the spectral weight distribution also reflects the behavior of the itinerant charges.  The development of a spin gap and the pile up of weight into a resonance peak are generally associated with cooling through $T_c$.

If the low-energy spin fluctuations are associated with particle-hole excitations near the Fermi level, then it is natural for a spin gap to develop when superconductivity occurs and gaps out the low-energy electronic states.\cite{lu92,bulu93}  Interactions can pull magnetic weight below twice the superconducting gap energy, resulting in the resonance peak.  This has been a very popular interpretation.\cite{esch06}

There are some results that the particle-hole excitation picture has difficulty explaining.  For example, in LSCO the resonance occurs at an incommensurate wave vector,\cite{tran04b,chri04} whereas the calculations tend to put it at \qaf.  Also, the calculated spectra for $T<T_c$ have the lower energy excitations dispersing in the directions along $(\frac12\pm\delta,\frac12\pm\delta)$ and $(\frac12\pm\delta,\frac12\mp\delta)$ (due to the nodal structure of the superconducting $d$-wave gap),\cite{lu92,zha93,norm07,breh10} which is inconsistent with experiment.\cite{lake99}

As we discuss in $\S$5.3, substituting large-moment Fe ions into overdoped cuprates can yield magnetically-ordered states that seem to require magnetic interactions via the conduction electrons.  Thus, it appears that conduction electrons will respond to embedded local moments.  The question in the underdoped regime is whether they can be responsible for magnetic correlations in the absence of any local moments.

From an alternative perspective, the low-energy spin excitations come from the same local Cu moments as the high energy ones.  Stripe correlations, as discussed in $\S$4, provide a motif in which mobile charge carriers and local magnetic moments can coexist.\cite{kive03}  The incommensurability of the low-energy excitations is then a direct result of the doped holes.  The development of a spin gap and resonance indicates a coherent response of the moments to the superconductivity, but a thorough explanation in terms of a stripe picture has yet to be provided.


\subsection{Evolution from spin waves to hour-glass}

\begin{figure}
\begin{center}
\includegraphics[width=8.2cm]{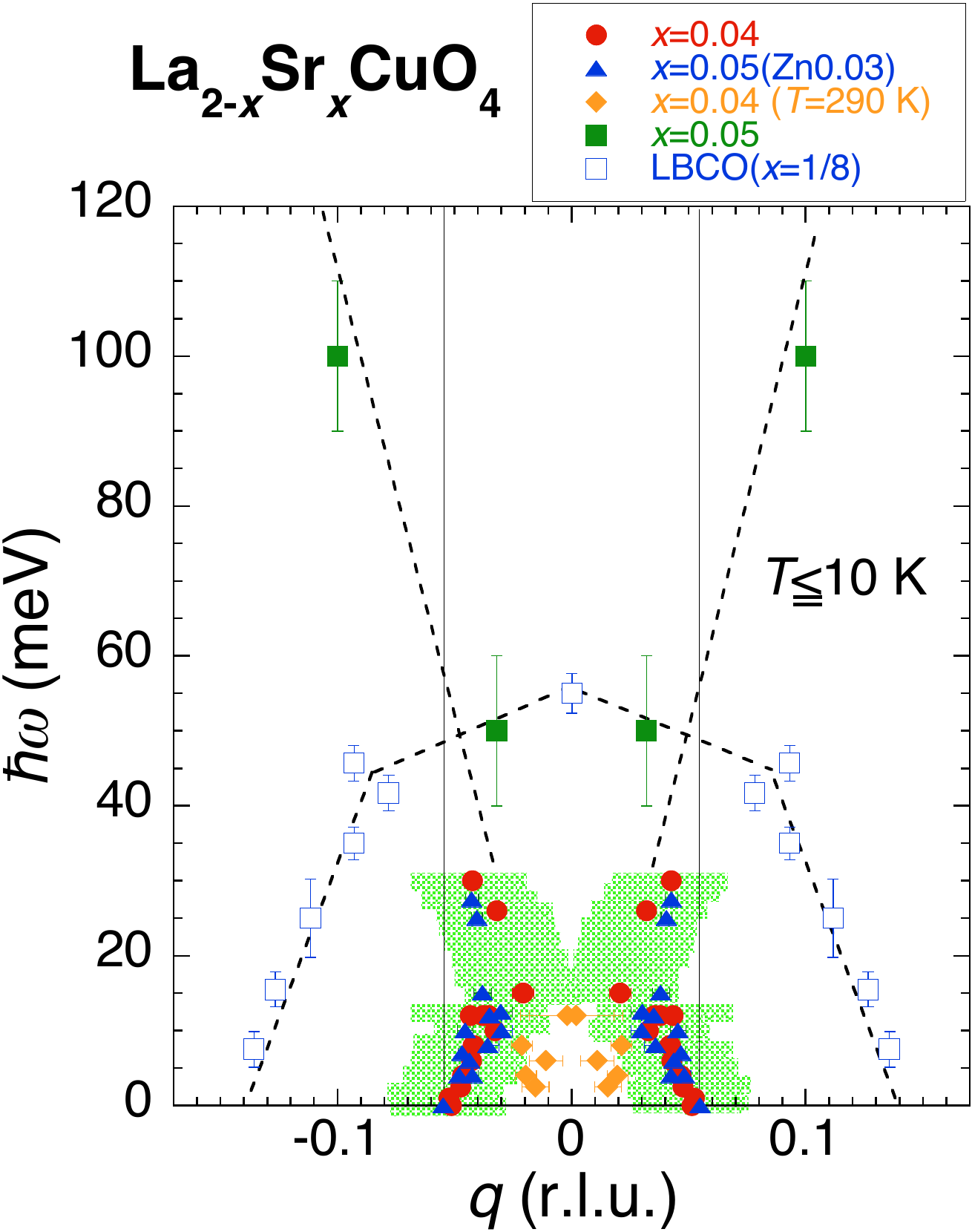}
\end{center}
\caption{(Color online) Magnetic dispersion relation along $q_K$ in La$_{1.96}$Sr$_{0.04}$CuO$_{4}$ (filled circles) below 10 K and at 290 K (filled diamonds) (Ref.~\cite{mats08}). For comparison, magnetic dispersion relations in other related compounds are also shown. The filled triangles, filled squares, and open squares are data of La$_{1.95}$Sr$_{0.05}$Cu$_{0.97}$Zn$_{0.03}$O$_{4}$, La$_{1.95}$Sr$_{0.05}$CuO$_{4}$ (Ref.~\cite{goka03}), and La$_{1.875}$Ba$_{0.125}$CuO$_{4}$ (Ref.~\cite{tran04}), respectively. It is noted that the peak positions are 45$^\circ$ rotated in La$_{1.875}$Ba$_{0.125}$CuO$_{4}$. The thick shaded bars represent the full width at half maximum of the excitation peaks in La$_{1.96}$Sr$_{0.04}$CuO$_{4}$. The broken lines are visual guides.
}
\label{fg:hg2}
\end{figure}

How does the magnetic dispersion change as one goes from the AF to the SC phase?  In LSCO, there is an intervening spin-glass regime, characterized by incommensurate spin order, with the ordering wave vectors rotated by 45$^{\circ}$ from those in the superconducting regime.\cite{birg06}  In orthorhombic notation, the spin ordering wave vectors are $(0,1\pm\delta,0)_{\rm o}$, with $\delta\sim x$.  Despite the rotation, the magnetic excitations exhibit a dispersion quite similar to the hour-glass of the superconducting regime.  Figure~\ref{fg:hg2} shows the dispersion determined for LSCO $x=0.04$, compared with that of a few related samples.\cite{mats08}

An interesting feature is that the low-energy spin fluctuations La$_{1.96}$Sr$_{0.04}$CuO$_4$ disperse only along the [010] direction and not along [100],\cite{mats08} similar to the 
nematic-like response of YBCO $x=0.3$, 0.35, and 0.45.\cite{hink08,haug10} This dynamical anisotropy is present even above the onset temperature of the static order. 
A theoretical study suggests that the nematic-like response originates from 
fluctuating spin stripes driven by the charge nematic ordering.\cite{sun10} 
The phenomenological model reproduces the observed magnetic excitation spectra.

\begin{figure}[t]
\begin{center}
\includegraphics[width=8.2cm]{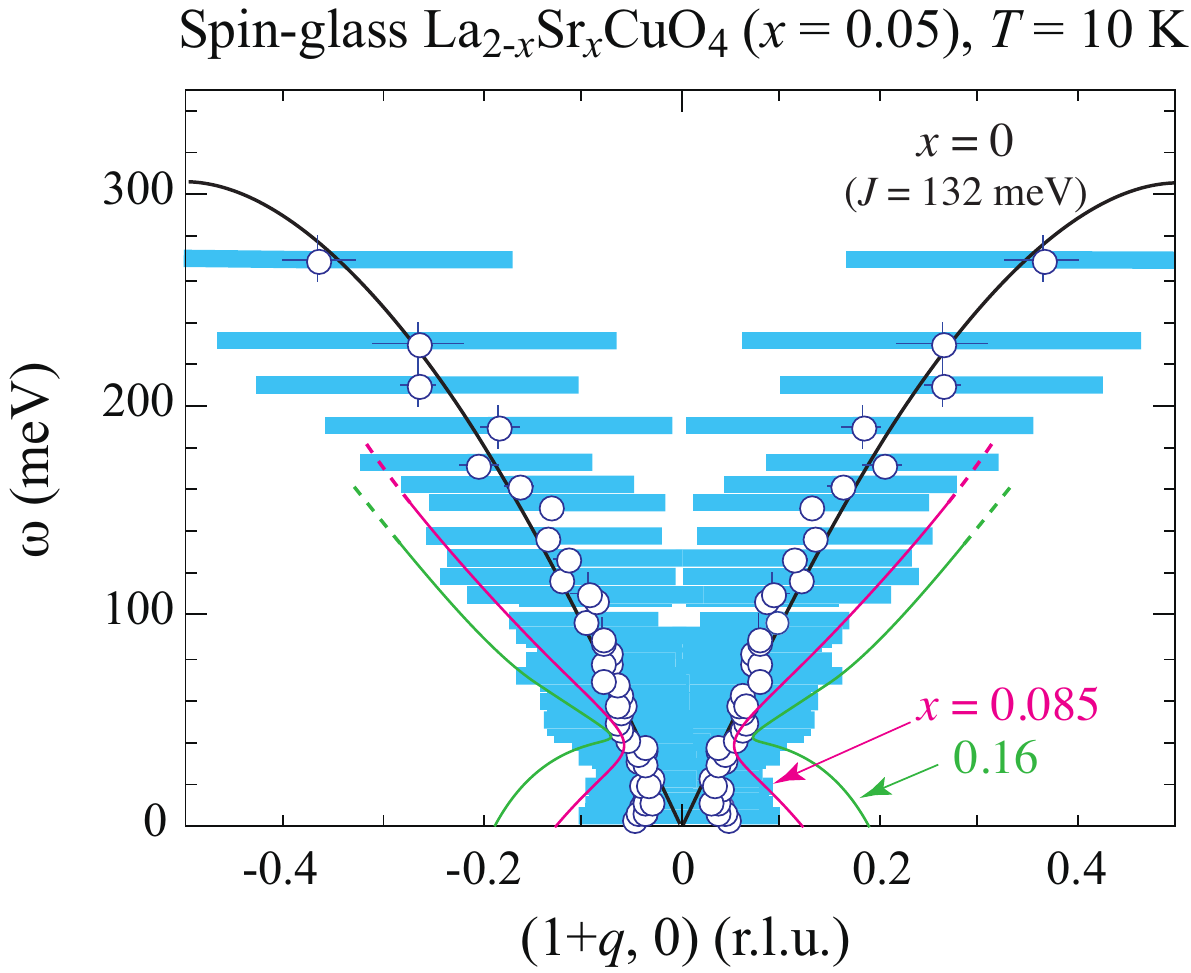}
\end{center}
    \caption{ (Color online)
    Magnetic dispersion of spin-glass LSCO with $x=0.05$ determined by time-of-flight spectroscopy.     The thick horizontal bars (light blue) represent peak width from constant-$\omega$ cuts.   For reference, results from LSCO with $x=0$ (antiferromagnetic insulator),  $0.085$ (underdoped superconductor), and $0.16$ (optimally-doped superconductor) are shown by separate (black, red, and green) curves, respectively.~\cite{hayd91,lips09,vign07} 
    }
    \label{fg:HHfig1}
\end{figure}

The excitations at energies above \ecross\ have been characterized by time-of-flight measurements on LSCO $x=0.05$.\cite{goka03}   The original data have been reanalyzed by Hiraka {\it et al.}\cite{hira11}, and the dispersion is presented in Fig.~\ref{fg:HHfig1}.  The high-energy dispersion is consistent with the spin waves of the $x=0$ phase, though with a slightly reduced $J$. Curves for LSCO $x=0.085$ and 0.16 are also shown in Fig.~\ref{fg:HHfig1}, and we can see that there is a gradual softening of the high-energy excitations, which can be described by a decrease in the effective $J$ describing the dispersion.  At the same time, the low-energy, long-wavelength excitations disperse from the incommensurate wave vectors.  Hole doping clearly reorganizes the excitations below \ecross, but has a more modest effect on the effective dispersion above it.

\begin{figure}
\begin{center}
\includegraphics[width=8.2cm]{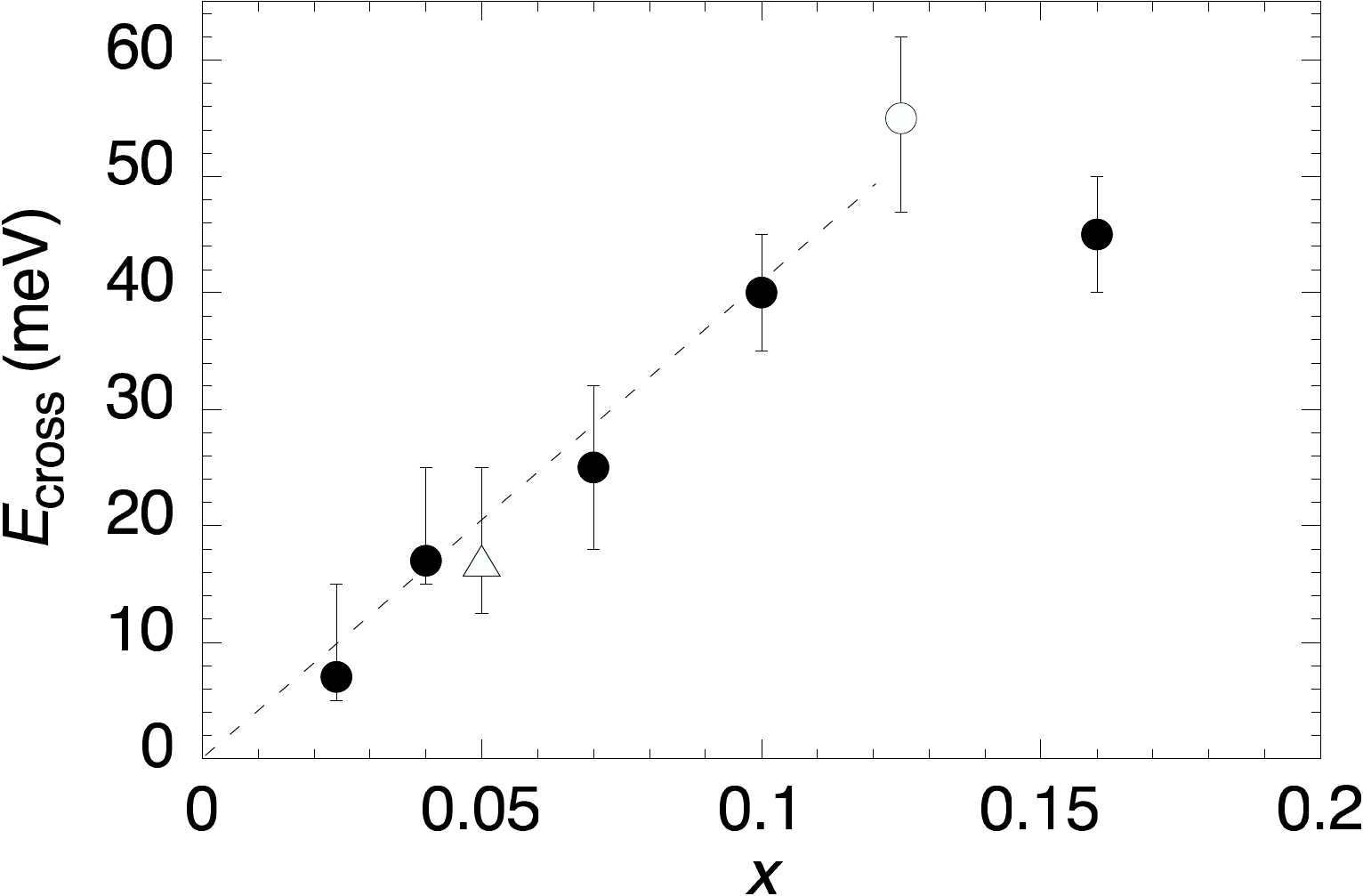}
\end{center}
\caption{Plot of \ecross\ vs.\ $x$ in LSCO (filled circles), La$_{2-x}$Sr$_x$Cu$_{1-y}$Zn$_y$O$_{4}$ (open triangle), and La$_{2-x}$Ba$_x$CuO$_4$ (open circle) (Refs.~\cite{mats00,hira01,kofu07,tran04,vign07,mats08}). The dashed line is a guide to the eye.}
\label{figEc}
\end{figure}

Returning to Fig.~\ref{fg:hg2}, it is intriguing that the slope of the effective low-energy dispersion, which corresponds to a velocity, shows little variation with doping.  If the velocity is independent of doping, while the incommensurability is linear in $x$ for $x\lesssim\frac18$, then one would expect \ecross\ to be proportional to $x$ \cite{bati01,krug03}. 
In Fig.~\ref{figEc}, values for \ecross\
extracted from neutron scattering studies \cite{mats00,hira01,kofu07,tran04,vign07} are plotted. Each symbol represents an estimate from interpolating a parabola through the low-energy dispersion, while the error bars indicate the energy range over which constant-energy cuts are consistent with a single peak of minimum width. The results are consistent with
$E_{\rm cross} \sim x$ for $x\lesssim\frac18$. 
This trend is opposite to the gradual decrease observed for high-energy, antiferromagnetic-like spin excitations,\cite{birg06} which we will discuss shortly.

Given the continuous evolution of \ecross\ with doping, providing a connection with La$_{1.875}$Ba$_{0.125}$CuO$_4$ where charge and spin stripe order is known to occur \cite{tran04}, it is considered that charge stripes and moment modulation are likely to be an important part of the incommensurate response in the diagonal incommensurate phase.
In terms of a stripe picture, the dominant magnetic interaction would be superexchange within locally antiferromagnetic domains. There is still a challenge to understand why the dispersion of the low-energy excitations is not significantly affected by the rotation in stripe orientation. One possibility suggested by Granath \cite{gran04} is that a diagonal stripe might consist of a staircase pattern of bond-parallel stripes, in which case local interactions would be independent of average stripe orientation. Granath\cite{gran04} found that such a pattern is necessary in order to obtain consistency with the photoemission experiments.\cite{yosh06} 

The magnetic excitations have been studied theoretically in a wide range of doping on the basis of the Hubbard model by Seibold and Lorenzana.\cite{seib06,seib09} The magnetic excitation spectra from the stripes almost reproduce the overall feature of the hour-glass excitations. However, the calculations predict much larger values for \ecross\ in the region of $x<$0.07 than those observed experimentally in LSCO, as shown in Fig.~\ref{fg:hg2}.
It has been suggested that much of that discrepancy can be eliminated by taking account of disorder, which softens and broadens the excitations.\cite{seib09}
Furthermore, it is noted that, while spin-wave calculations for a diagonal-stripe model \cite{krug03,carl04} provide a good description of the magnetic spectrum observed in insulating La$_{2-x}$Sr$_x$NiO$_4$,\cite{bour03,woo05} they have difficulty in reproducing the hour-glass-like spectrum in LSCO.

\subsection{Magnetic spectral weight}

Besides the dispersion, it is also important to consider how the frequency-dependent magnetic spectral weight evolves with doping.  Neutron scattering directly measures the dynamical structure factor, ${\cal S}({\bf Q},\omega)$, which is proportional to $\chi''({\bf Q},\omega)/(1-e^{-\hbar\omega/k_{\rm B}T})$, where $\chi''({\bf Q},\omega)$ is the imaginary part of the dynamical susceptibility.  By integrating $\chi''({\bf Q},\omega)$ over {\bf Q} (within a Brillouin zone) one obtains the local susceptibility, $\chi''(\omega)$.   There is a sum rule for ${\cal S}({\bf Q},\omega)$; integrating ${\cal S}$ over {\bf Q} and $\omega$ yields $\langle {\bf S}^2\rangle$, corresponding in the present case to the mean-squared spin per Cu atom.  The thermal factor connecting ${\cal S}$ and $\chi''$ goes to unity as $T\rightarrow0$, so integrating the low-temperature local susceptibility over $\omega$ also gives a measure of the mean-squared spin.

\begin{figure}[t]
\begin{center}
\includegraphics[width=8.2cm]{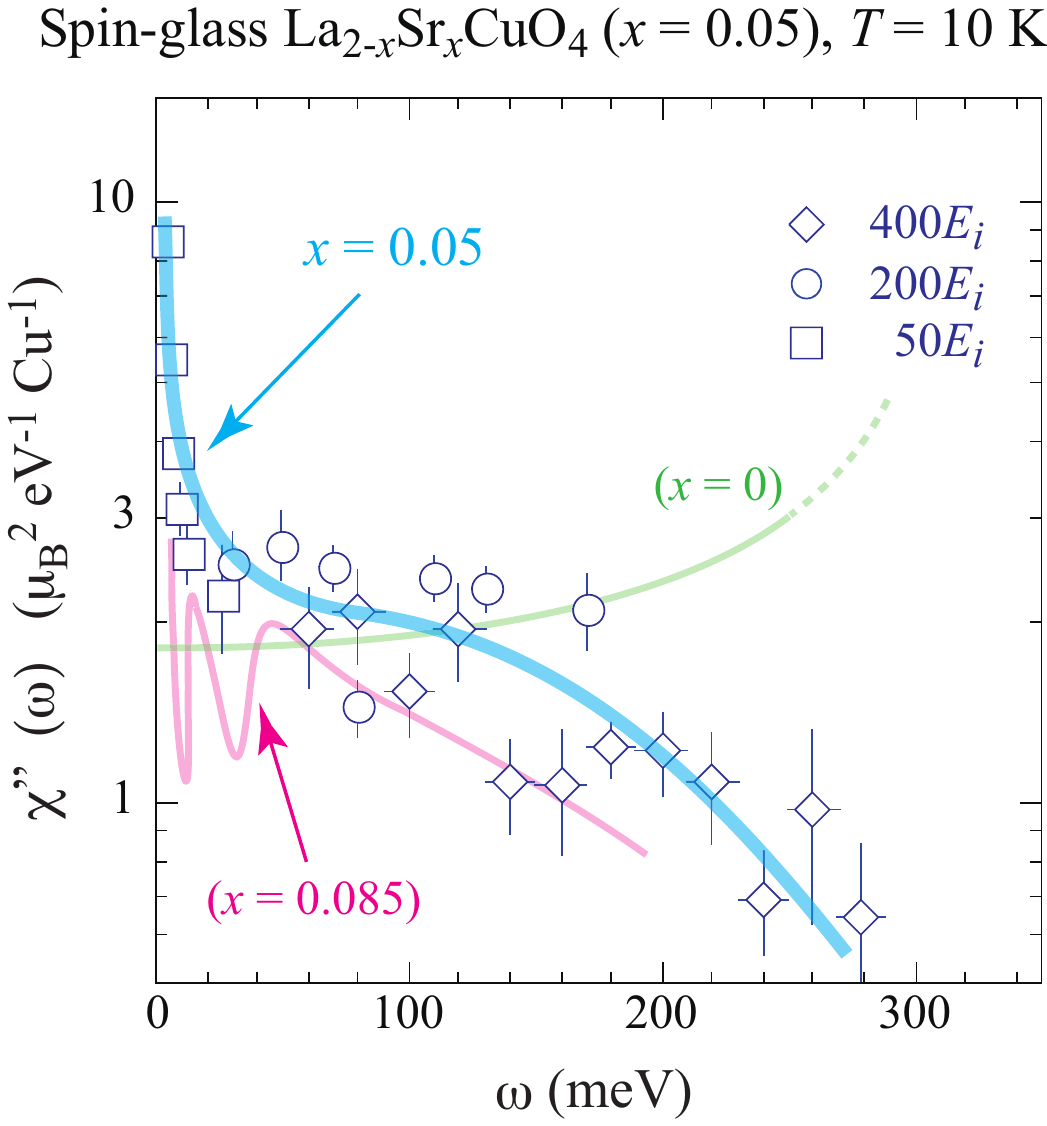}
\end{center}
    \caption{(Color online)
$\chi^{\prime\prime}(\omega)$ at 10~K determined by pulsed-neutron scattering experiments.
The line (light-blue) through the data points is a guide to eyes.
Results from LSCO with $x=0$ and $0.085$ are indicated by solid lines (green and red, respectively).~\cite{hayd91,lips09}   
}
    \label{fg:HHfig2}
\end{figure}

We begin by considering $\chi''(\omega)$ in LSCO $x=0.05$, as shown in Fig.~\ref{fg:HHfig2}.\cite{goka03,hira11}  In the range of 50 to 150~meV, the magnitude is comparable to that for spin waves in the ordered antiferromagnet at $x=0$.  There is a substantial upturn at low frequency, which appears to correspond to the weight that correspond to the static order parameter in the AF.  Thus, in the spin-glass phase, the frustration of AF order by the doped holes causes the low-energy correlations to remain dynamic.  Comparing with the results for LSCO $x=0.085$,\cite{lips09} indicated by the red curve in Fig.~\ref{fg:HHfig2}, we see that the enhanced low-energy weight appears to be suppressed as one moves into the superconducting regime.

\begin{figure}[t]
\begin{center}
\includegraphics[width=8.2cm]{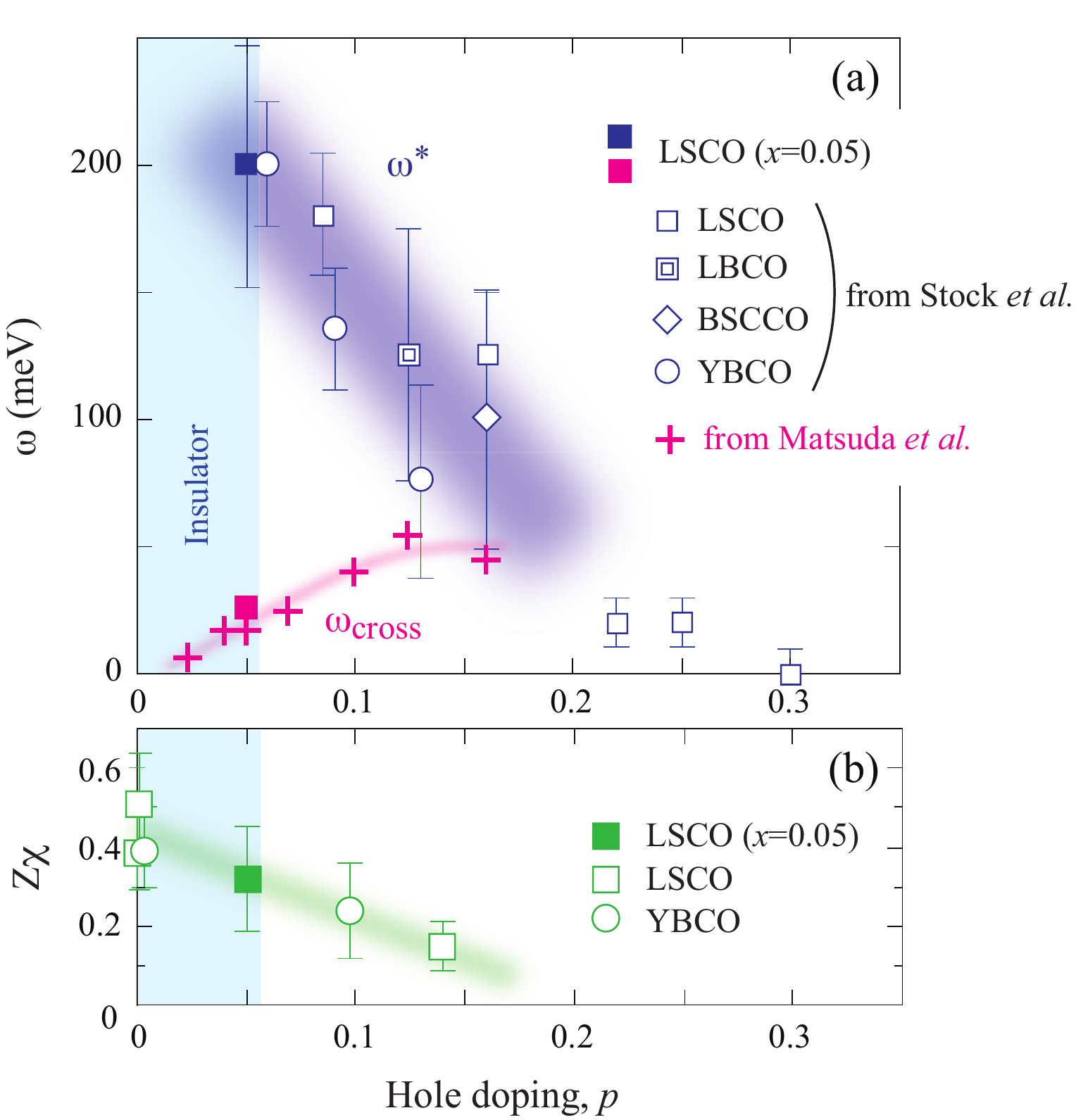}
\end{center}
\caption{(Color online)
	(a) $\omega^{\ast}$ plot following Stock \textit{et al.}\cite{stoc10}, where open symbols indicate for a number of cuprates the energy at which $\chi''(\omega)$ becomes half of that of undoped La$_2$CuO$_4$. Stock \textit{et al.}\cite{stoc10} have pointed out that the $p$ dependence $\hbar\omega^\ast$ is quite similar to that of $\hbar\omega_{\rm pg}$ determined by various electronic probes.\cite{hufn08}  The crosses indicate \ecross, from Fig.~\ref{figEc}.  The filled symbols are the corresponding points for LSCO $x=0.05$.\cite{hira11}  (b) Plot of $Z_\chi$ vs.\ hole doping for LSCO, where $Z_\chi$ is the ratio of the frequency-integrated local susceptibility to that predicted by spin-wave theory for a 2D AF.\cite{hira11}  Filled symbol is for LSCO $x=0.05$; open squares, LSCO\cite{cold01,hayd96a}; open circles, YBCO.\cite{hayd96b}}
\label{fg:xdep}
\end{figure}

We can also see from Fig.~\ref{fg:HHfig2} that $\chi''(\omega)$ for $x=0.05$ begins to fall substantially below the $x=0$ result for $\hbar\omega\gtrsim200$~meV.  For $x=0.085$, the fall off begins at a slightly lower energy.  Stock {\it et al.}\cite{stoc10} were the first to identify this trend, and to show that it occurs in several different cuprate families.  In particular, they estimated the energy $\hbar\omega^\ast$ at which $\chi''(\omega)$ falls to half of that for that AF phase.  In the top of Fig.~\ref{fg:xdep}, we have reproduced their plot, together with a point for the LSCO $x=0.05$ sample.  Further evidence for a universal trend of reduction in magnetic bandwidth with doping is provided by Raman measurements of two-magnon scattering.\cite{suga03}  Stock {\it et al.}\cite{stoc10} also pointed out that this energy scale is quite similar to the pseudogap energy that has been determined from a number of electronic probes, such as ARPES and STS.\cite{hufn08}  We note that this energy scale is also virtually identical to the energy gap determined from an analysis\cite{gork06} of the temperature dependence of the Hall effect in LSCO,\cite{ando04} and it has the same doping dependence as the mid-infrared gap determined by optical reflectivity measurements on YBCO.\cite{lee05}

Why would magnetic spectral weight disappear above the pseudogap energy?  Consider first the parent insulator phase.  The AF order and spin-waves are well-defined there because they occur at energies well below the gap ($\sim2$~eV) for charge excitations.\cite{kast98}  The superexchange energy that drives the AF correlations is a consequence of the competition between the strong onsite Coulomb repulsion between Cu $3d$ electrons and the kinetic energy of these electrons, which can be reduced by hopping between neighboring sites.\cite{ande97}  For this magnetic mechanism to survive hole doping, it should be favorable to maintain a particle-hole excitation gap.  When magnetic excitations exceed that gap, they may no longer be defined, a connection noted by Stock {\it et al.}\cite{stoc10}  It has been established from ARPES studies\cite{dama03} that the pseudogap has a strong dependence on the electronic momentum {\bf k}.  For {\bf k} oriented at $\sim45^\circ$ to the Cu-O bonds (nodal direction), there is no gap, but the pseudogap is large for {\bf k} parallel to the Cu-O bonds (antinodal direction).  If we think about charges in an AF background,\cite{trug88,lau11} then an electron near the Fermi energy moving in the nodal direction can hop on the same AF sublattice; no spin flips are involved, so there is no conflict with the AF correlations.  In contrast, an electron hopping in the antinodal direction, along Cu-O bonds, can only do so by flipping spins and disrupting the AF correlations.  Thus, it is physically reasonable that the antinodal pseudogap sets an upper limit for the existence of locally-AF spin correlations.

There is also interesting structure and temperature dependence in the spectral weight for $\hbar\omega\lesssim E_{\rm cross}$.  As mentioned in \S\ref{sc:back}, near optimal doping and above, a spin gap opens below $T_c$ and weight moves into a resonance peak above it.  In YBCO and Bi2212, the resonance energy is similar to \ecross.  In LSCO, the spin gap is much smaller than \ecross; for optimal doping, the resonance energy is $\sim18$~meV,\cite{vign07,chri04} and there is still a feature there just above $T_c$.\cite{chri04}  Lipscombe {\it et al.}\cite{lips09} have followed the temperature evolution of the the structures in $\chi''(\omega)$ below \ecross\ on cooling from 300~K to low temperature.  We have plotted the doping dependence of \ecross\ for the the LSCO-related samples with red crosses in Fig.~\ref{fg:xdep} to contrast it with $\hbar\omega^\ast$.   It is comparable to the upper limits for $2\Delta_{\rm sc}$, where $\Delta_{\rm sc}$ is the effective superconducting gap determined at the edge of the Fermi arc, as identified in a recent ARPES study.\cite{idet11}

\begin{figure}[t]
\begin{center}
\includegraphics[width=7.7cm]{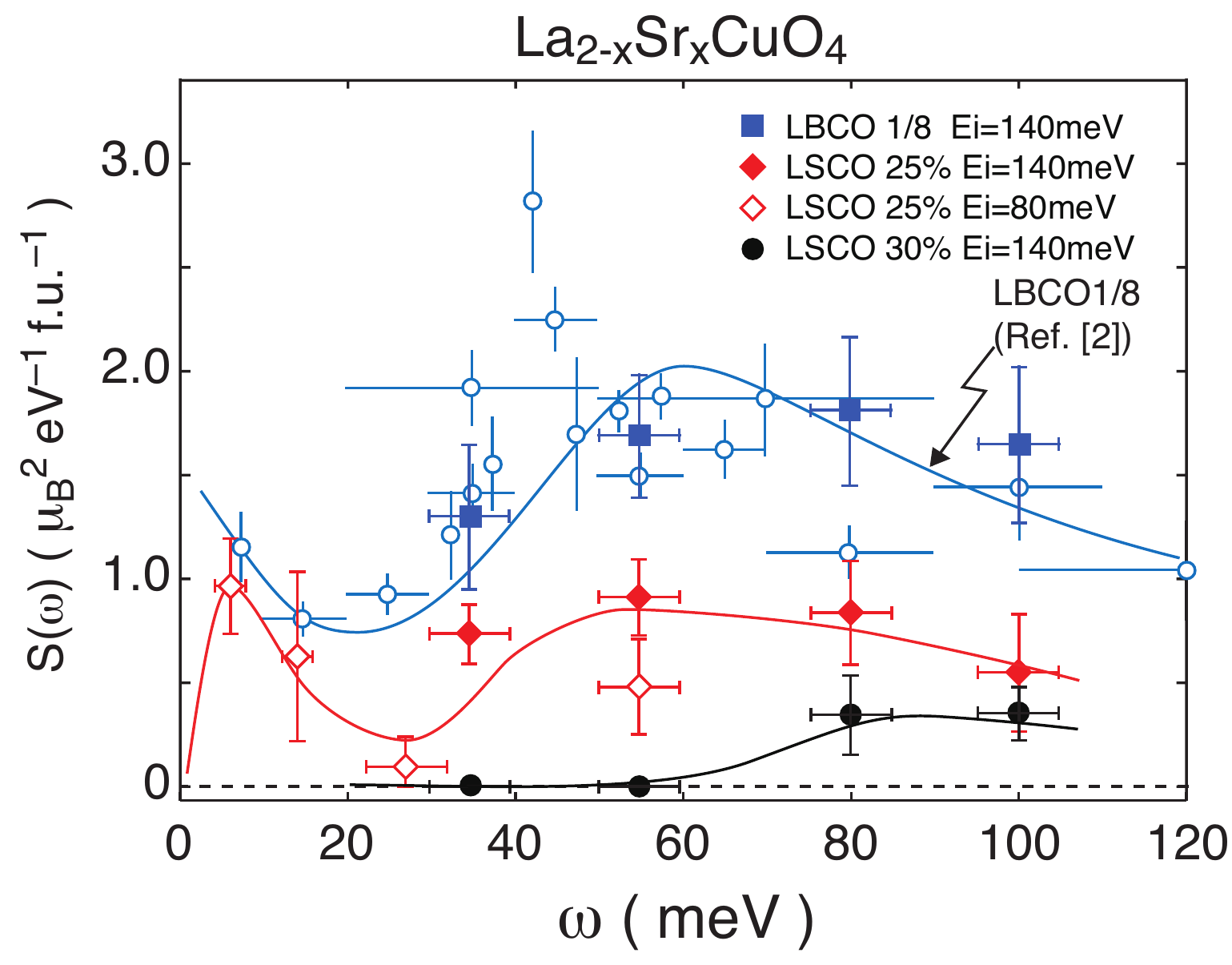}
\end{center}
\caption{(Color online)  $S(\omega)$ spectra of overdoped La$_{2-x}$Sr$_{x}$CuO$_{4}$ with $x=0.25$ and 0.30 compared with La$_{2-x}$Ba$_{x}$CuO$_{4}$ with $x=1/8$, from Ref.~\cite{waki07b}.}
\label{fg:waki1}
\end{figure}

The overdoped regime provides another opportunity to probe the relationship between spin fluctuations and superconductivity.  The evidence for depressed magnetic spectral weight at low frequencies has been discussed previously.\cite{birg06}  More recently, Wakimoto {\it et al.}\cite{waki07b} have measured the magnetic excitations of La$_{2-x}$Sr$_{x}$CuO$_{4}$ (LSCO) with $x=0.25$ and 0.30 up to 100 meV using pulsed neutrons at ISIS.  They found that $\chi''(\omega)$ is diminished over the entire energy range compared to underdoped samples.  In particular, magnetic excitations at $\omega<60$~meV have completely vanished in the non-superconducting $x=0.30$ sample, as shown in Fig.~\ref{fg:waki1}.  In related work, Lipscombe {\it et al.}\cite{lips07} reported a qualitatively similar spin excitation spectrum in the overdoped $x=0.22$ sample, with a strong depression of the local susceptibility in the range of 40--70 meV.  These results are consistent with a gradual suppression of AF spectral weight with overdoping, and in parallel with the reduction in $T_c$.  
	
The bottom panel of Fig.~\ref{fg:xdep} shows $Z_\chi$, the ratio of the frequency-integrated $\chi''(\omega)$ relative to the spin-wave theory prediction for a two-dimensional AF, obtained for LSCO.  It demonstrates that doping causes a gradual decrease of the mean-square magnetic moment as one moves away from the AF phase.  There are experimental indications for a similar trend in YBCO.\cite{regn95,kive03,rezn08}  Therefore it is likely to be a universal trend in the hole-doped cuprate superconductors that the antiferromagnetic spectral weight decreases with doping, especially in the overdoped regime.  We note, however, that a recent resonant inelastic x-ray scattering (RIXS) study of spin fluctuations in YBCO has interpreted the measurements as indicating rather little renormalization of $J$ or reduction of spectral weight with doping.\cite{leta11}  Further comparisons of RIXS and neutron measurements on similar samples are needed to resolve this discrepancy.
	
Uemura\cite{uemu03} has reviewed evidence that a reduced fraction of the normal-state carriers participate in the superfluid density for overdoped samples.  He has argued that there may be a short length scale phase separation between normal and superconducting regions.  If AF spin correlations are important to the superconductivity, such a phase separation and the reduced superfluid density in overdoped samples would be compatible with the reduced magnetic spectral weight.

\section{Stripes and superconductivity}
\label{sc:stripes}

\subsection{Impact of crystal symmetry}

The downward-dispersing excitations of the hour-glass spectrum connect (in the case of LSCO and LBCO), or extrapolate (in the case of YBCO), to the incommensurate wave vectors associated with spin-stripe order.  For 214 cuprates with a lattice symmetry that makes orthogonal Cu-O bonds inequivalent, both charge and spin stripe order have been experimentally identified.\cite{tran95a,fuji04}  The role of charge and spin stripes has been controversial.\cite{kive03,tran07,vojt09}  Much of the focus has been in terms of a type of order that competes with superconductivity; however, recent observations of two-dimensional (2D) superconducting correlations coexisting with stripe order\cite{li07,tran08} have led to suggestions of a more intimate connection with pairing and the phase of the superconducting order parameter.\cite{hime02,berg07,berg09b}

\begin{figure}[t]
\begin{center}
\includegraphics[width=8.2cm]{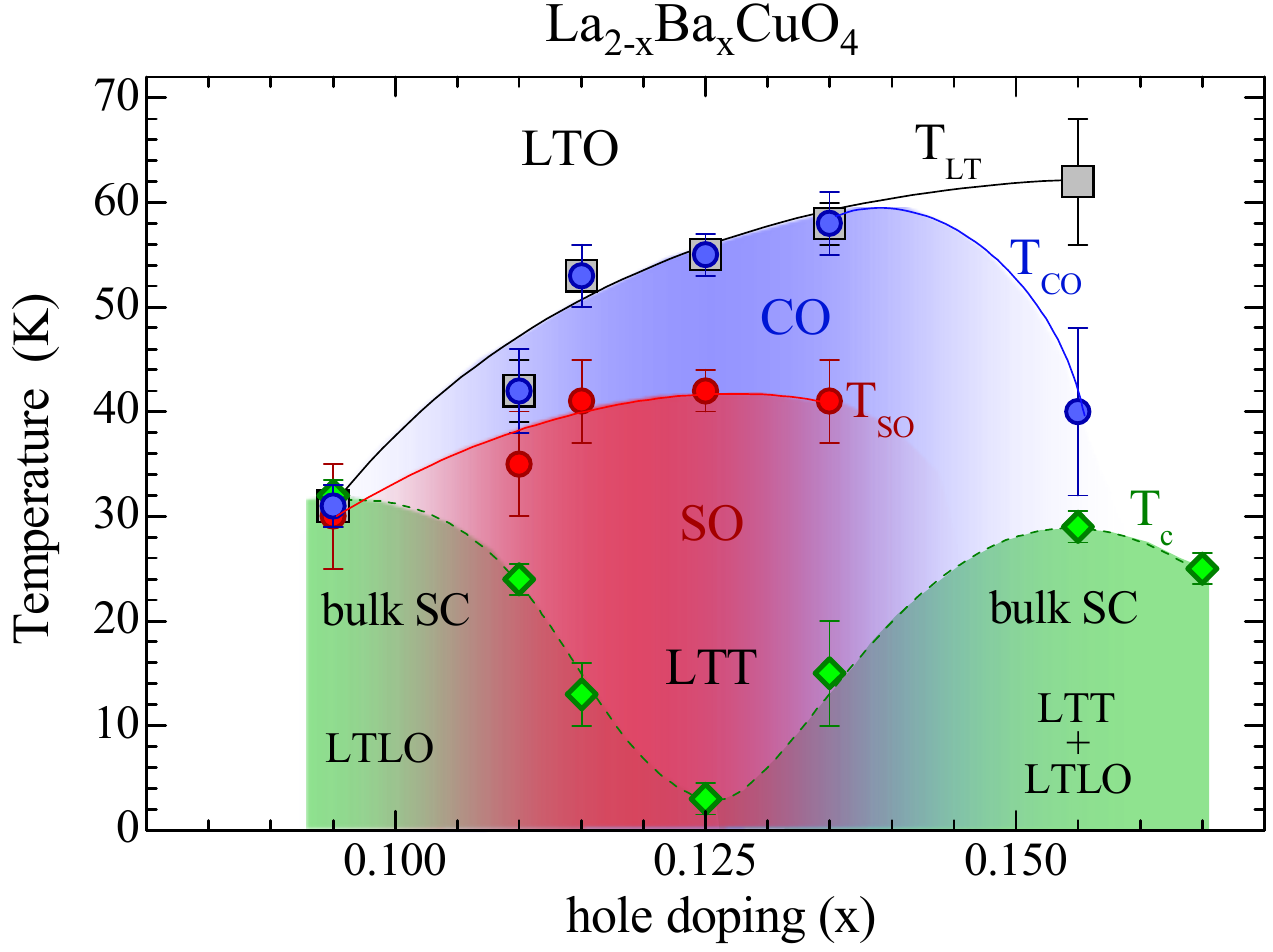}
\end{center}
\caption{(Color online)  Phase diagram for \lbco\ as a function of temperature and doping as determined from single crystals.\cite{huck11}    Transitions are indicated as follows: structural transition, $T_{\rm LT}$, (gray) squares; charge stripe order (CO), $T_{\rm CO}$, (blue) circles; spin stripe order (SO), $T_{\rm SO}$, (red) circles; bulk superconducting $T_c$, (green) diamonds.  The low-temperature phase is either low-temperature tetragonal (LTT), low-temperature less orthorhombic (LTLO), or a coexistence of the two.}
\label{fg:lbco}
\end{figure}

Several groups have explored the doping dependence of stripe order in \lbco\ with neutron and x-ray scattering.\cite{fuji06,kim08,duns08,duns08b,huck11}  A phase diagram for stripe order in \lbco, as reported by H\"ucker {\it et al.},\cite{huck11} is shown in Fig.~\ref{fg:lbco}.  As one can see, the onset of charge stripe order is limited by the structural transition from the low-temperature orthorhombic (LTO) phase to the low-temperature tetragonal (LTT); static spin stripe order develops at a lower temperature.   When comparing the results from various studies, there are some quantitative discrepancies regarding stripe ordering temperatures at particular doping levels; however, these are likely due to uncertainties in the Ba concentration.  It has been shown\cite{huck11} that the discrepancies can be resolved by calibrating the composition through the doping-dependent transition temperature from the high-temperature tetragonal phase (HTT) to the LTO,\cite{zhao07} with the assumption that the transition temperature varies linearly with $x$.  Note that Dunsiger {\it et al.}\cite{duns08b} have also confirmed the rotation in stripe direction from vertical to diagonal on reducing $x$ from 0.08 to 0.05 and 0.025.

As demonstrated by Abbamonte {\it et al.},\cite{abba05} charge stripe order can also be detected by  resonant soft x-ray diffraction.  That technique has now been used by Fink {\it et al.}\cite{fink11} to determine the phase diagram for charge order in \lesco.  The Eu-doped system is of interest because the charge-ordering transition is well separated from the structural transition, from LTO to LTT, whereas the structural transition appears to limit the onset of charge stripe order in \lbco, as indicated in Fig.~\ref{fg:lbco}.  The occurrence of spin stripe order in \lesco\ was previously determined by muon spin-rotation spectroscopy\cite{klau00} and confirmed for one composition ($x=0.15$) by neutron diffraction.\cite{huck07}

	An early observation of checkerboard-like modulations in the electronic density of states for \bscco\ and in \nccoc\ observed by scanning tunneling spectroscopy (STS)\cite{hoff02,hana04} caused some researchers to raise questions\cite{vojt02,fine04,sush04} about the interpretation of the spin and charge order peaks detected previously by neutron and x-ray diffraction.  As a check, Christensen {\it et al.}\cite{chri07} used polarized neutrons to characterize the magnetic superlattice peaks and the low-energy spin fluctuations in a crystal of La$_{1.48}$Nd$_{0.4}$Sr$_{0.12}$CuO$_4$.  Their results were consistent with a unidirectional stripe modulation together with collinear spin order, although they could not rule out a more complicated two-dimensionally modulated noncollinear spin structure.  It is interesting to note that more recent STS studies, especially on \bscco, have found that the electronic modulations break four-fold rotational symmetry, consistent with a locally unidirectional modulation.\cite{kohs07,lawl10}  In fact, Parker {\it et al.}\cite{park10} have shown that stripe-like modulations in \bscco\ are strongest for hole concentrations $p\sim 1/8$, similar to the 214 cuprates.

\begin{figure}[t]
\begin{center}
\includegraphics[width=6.9cm]{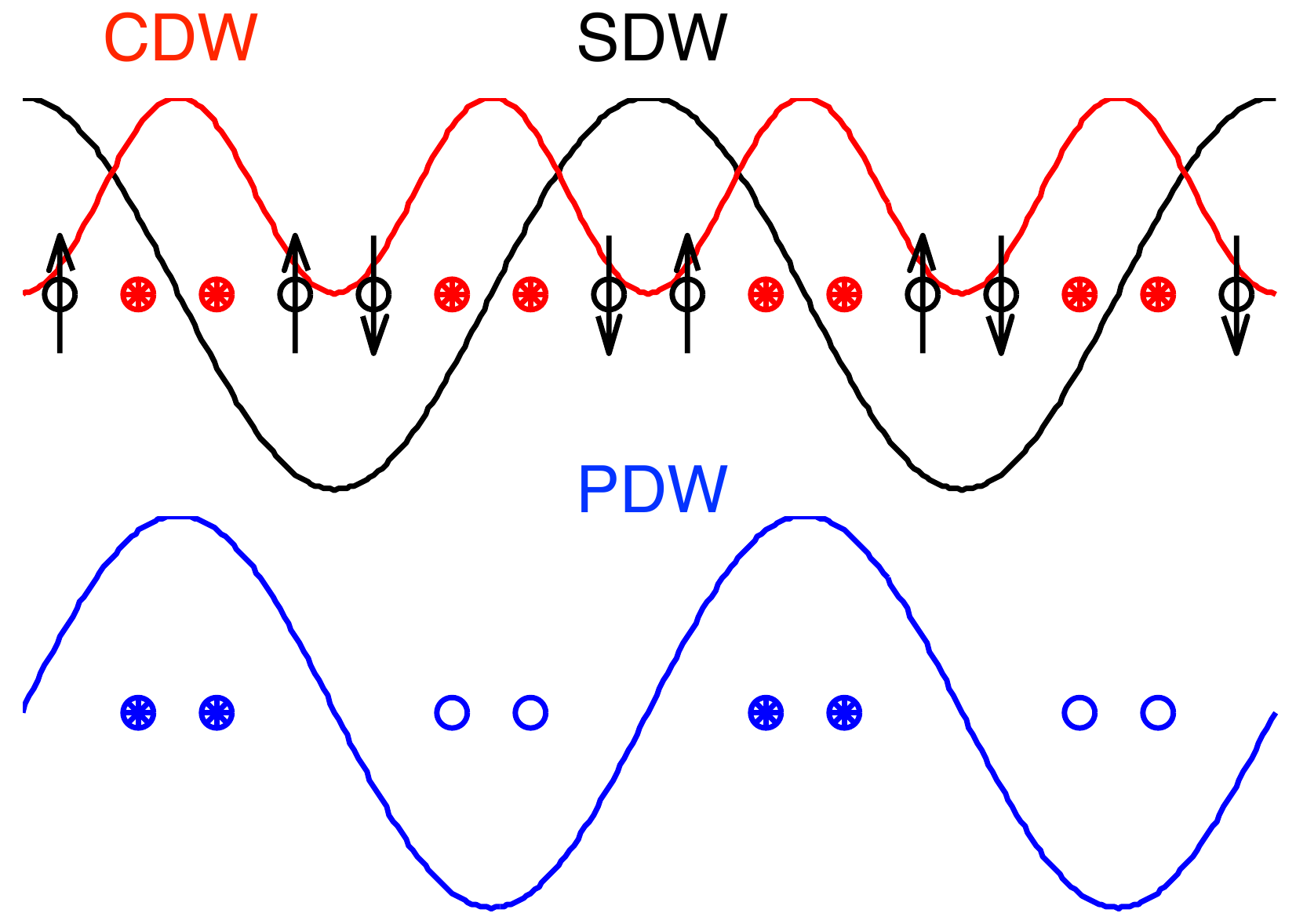}
\end{center}
\caption{(Color online)  Schematic diagram of CDW, SDW, and PDW orders, indicating the relationships among the phases of the modulations.}
\label{fg:pdw}
\end{figure}

As mentioned above, the detection\cite{li07,tran08} of 2D superconducting correlations, appearing concomitantly with spin stripe order, in \lbco\ with $x=1/8$ has changed the discussion concerning the nature of stripes.  The superconductivity in the cuprates is driven by interactions within the CuO$_2$ planes; however, the presence of a Josephson coupling between the planes drives the superconductivity to 3D order before 2D order can be detected.  To explain the results in LBCO, it is necessary to find a mechanism associated with the stripe order that can frustrate the interlayer Josephson coupling.  The concept of the pair-density-wave (PDW) superconductor has been proposed as part of one such mechanism.\cite{hime02,berg07}  The idea is that the pair wave function has $d$-wave character within each charge stripe, but, in contrast to the uniform $d$-wave state, the phase of the pair wave function changes phase by $\pi$ from one stripe to the next, as indicated in Fig.~\ref{fg:pdw}.  When one combines this finite {\bf q} superconductivity with the fact that the stripe orientation rotates $90^\circ$ from one layer to the next in the LTT phase, it is not hard to see that the Josephson coupling should average to zero (for 3D long-range stripe order).  The PDW state was originally introduced\cite{hime02} to explain the decrease in the Josephson plasma resonance detected by $c$-axis infrared reflectivity\cite{taji01} on approaching the LTT phase, and the onset of stripe order, in La$_{1.85-y}$Nd$_y$Sr$_{0.15}$CuO$_4$.

In the case of \lbco, the PDW concept was initially expected to explain both the frustration of the interlayer Josephson coupling and the onset of strong 2D superconducting correlations\cite{berg07}; however, new experiments have led to a more complicated interpretation.\cite{berg09b}  Analysis of the gap structure of the PDW state\cite{baru08} indicates that there should be a large gap in the antinodal electronic states, but that the nodal arc should be gapless.  In principle, a $d$-wave gap can develop on the nodal arc associated with a uniform $d$-wave state.  Angle-resolved photoemission studies\cite{vall06,he09} on LBCO $x=1/8$ have found evidence for a $d$-wave-like gap at low temperature, with the near-nodal gap closing near 40~K,\cite{he09} where the 2D superconducting correlations also disappear.  A possible scenario is that PDW correlations develop together with the charge stripe order, and that uniform $d$ wave superconductivity develops on top of this below 40~K.

Stripe-like spin order is not unique to the 214 cuprates.  Incommensurate elastic peaks have also been observed in \ybco\ with $x=0.3$, 0.35, and 0.45 (corresponding to $p=0.052$, 0.062, 0.082, and $T_c=0$, 10~K, and 35 K, respectively) by Hinkov and coworkers.\cite{hink08,haug10}  The ability to resolve the split magnetic peaks was enabled by the use of detwinned crystals.  (Stock {\it et al.}\cite{stoc08} did not resolve incommensurability in the elastic magnetic scattering from an $x=0.35$, $T_c=18$~K, twinned crystal.)  In fact, it was demonstrated that the spin modulation direction is uniquely oriented with respect to the Cu-O chains; in particular, the spin and chain modulation directions are parallel.   Furthermore, Suchanek {\it et al.}\cite{such10} found that doping YBCO $x=0.6$ with 2\%\ Zn shifted magnetic weight from the resonance to low-energy incommensurate peaks.  The magnetic incommensurability in YBCO increases with doping, but $\delta$ is significantly smaller than $p$, in contrast to the behavior in 214 cuprates.\cite{haug10}

The elastic magnetic peaks are only observed below $T_{\rm SDW}\sim 40$~K; however, the incommensurability in low-energy magnetic excitations can be resolved up to $\sim 150$~K.\cite{hink08,haug10}  The onset of the anisotropy in the spin dynamics has been discussed in terms of the development of nematic electronic correlations.\cite{kive98}  The onset temperature is comparable to that for in-plane anisotropy of the Nernst effect,\cite{daou10}  and also to the onset of bilayer superconducting correlations identified by $c$-axis optical conductivity.\cite{dubr11}

The elastic magnetic signal in YBCO disappears for $x\gtrsim0.5$, as a spin gap opens; nevertheless, a ``1/8'' effect has been identified by Taillefer and coworkers by examining transport properties in high magnetic fields.  For example, they have shown\cite{lali11} that, when superconductivity is suppressed by a $c$-axis magnetic field, the temperature at which the thermopower changes sign is maximum at $p=1/8$ both in YBCO and in \lesco.  Similarly, the temperature at which the Hall constant changes sign in YBCO has a maximum at that same point.\cite{lebo11}

Boothroyd {\it et al.}\cite{boot11} have recently reported a study of spin excitations in an insulating, stripe-ordered cobaltate system, La$_{5/3}$Sr$_{1/3}$CoO$_4$.  The interesting feature here is that the magnetic dispersion has the hour-glass shape of the cuprates.  This example clearly demonstrates that such a dispersion can occur in the absence of itinerant electrons.

\subsection{Effect of a magnetic field}

Neutron scattering studies of magnetic-field effects on the incommensurate spin density wave (SDW) order from the stripe phase, as well as on the dynamical spin excitations, have been performed mostly using the LSCO system.  This system exhibits the SDW order in the underdoped superconducting regime ($0.06<x<0.14$) while the system has a gap in the spin excitations in the optimally and slightly overdoped regime ($0.14<x<0.22$).  In this section we review the experimental results by categorizing the magnetic field effects on the SDW state in the underdoped samples and on the spin excitations in the spin-gapped, optimally-doped samples.

Pioneering work was done by Katano {\it et al.}\cite{kata00} who reported a small enhancement of the incommensurate magnetic order in LSCO with $x=0.12$ by application of magnetic field of 10 T along the $c$-axis.  More drastic enhancement by magnetic fields were observed in underdoped LSCO\cite{lake02,chan08} and stage-4, 6 La$_{2}$CuO$_{4+y}$.\cite{khay02,khay03}  The Ba-doped system LBCO is known to have a robust stripe order and a limited superconducting phase near 1/8 doping associated with the LTT structure.  The magnetic field effect on the stripe order was reported to be very limited in underdoped LBCO\cite{duns08} and in LBCO with $x=1/8$.\cite{wen08b}  Nd-doped La$_{2-x}$Sr$_{x}$CuO$_{4}$ also has a well-developed stripe order.  In this case, a $c$-axis magnetic field suppresses the subordinate order of Nd spins, but the stripe order itself is not affected by magnetic fields up to 4 T.\cite{waki03}  These results indicate that the magnetic field generally enhances the SDW state at a cost of the superconducting volume faction, but the degree of enhancement depends on the volume fraction of SDW and superconducting phases in zero field.  This behavior suggests that the ``normal'' state achieved by the suppression of superconductivity in magnetic vortex cores involves stripe order.\cite{deml01,kive02} 

The magnetic field effect on dynamical spin fluctuations was first reported by Lake {\it et al.}\cite{lake01} They found that application of a 7.5-T magnetic field on the spin-gapped, optimally-doped LSCO induces an additional spin fluctuation spectrum below the gap energy.  Further detailed measurements have shown the redistribution of the spectral weight from above to below the gap energy.\cite{gila04,tran04b,chan09}

\begin{figure}[t]
\begin{center}
\includegraphics[width=8.2cm]{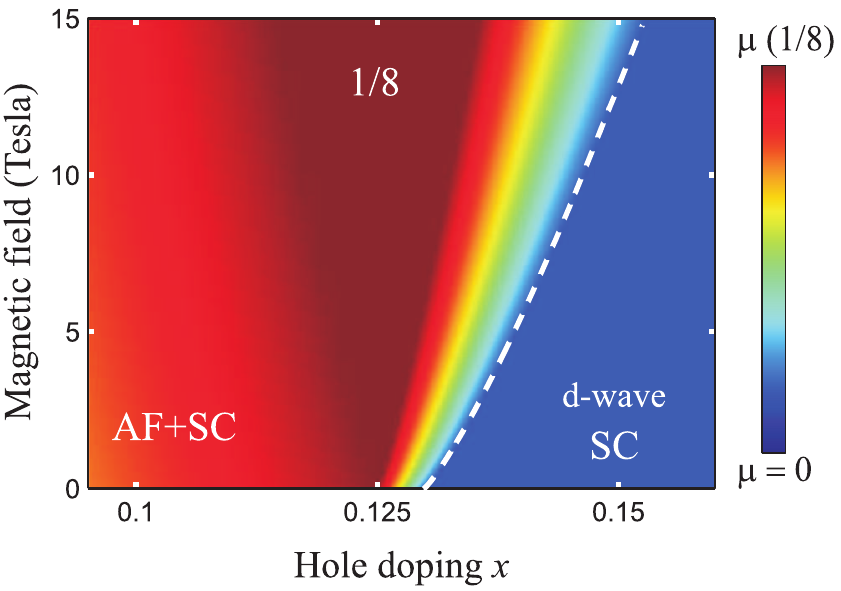}
\end{center}
\caption{(Color online)  Phase diagram indicating relative magnitude of ordered magnetic moment as a function of magnetic field and hole doping in LSCO.  Reprinted with permission from Chang {\it et al.}\cite{chan08} \copyright 2008 by the American Physical Society.}
\label{fg:field}
\end{figure}

The study of magnetic field effects can provide a test of the theoretical prediction of a quantum phase transition between the superconducting phase and one involving the coexistence of SDW and superconducting orders.\cite{deml01,kive02}  Khaykovich {\it et al.}\cite{khay05} performed a detailed study of LSCO samples with $x$ between 0.14 and 0.15, evenutally finding that a finite magnetic field induces SDW order in the $x=0.144$ sample, which is pure superconducting phase, with a small spin gap, in zero field.  This is direct evidence of the quantum phase transition, and the composition $x=0.144$ is in vicinity of the quantum critical point.  Later, Chang {\it et al.}\cite{chan08} have observed the magnetic-field-induced SDW order for $x=0.145$ above $\sim7$~T.  Accordingly, the spin gap decreases with magnetic field and disappears at $\sim7$~T.\cite{chan09} 
Combined with the results of muon spin rotation ($\mu$SR) studies, they have summarized the situation with the schematic doping-field phase diagram shown in Fig.~\ref{fg:field}.

Recently, in a study of LBCO with $x=0.095$, a field-induced enhancement for charge stripe order was observed, along with the enhancement of spin order.\cite{wen11}  Weak charge order is generally more difficult to detect than spin order, which might explain why this effect was not detected before.  Given the strong connection between charge and stripe order, there is a good possibility that the field-induced SDW results also have an associated charge order.

The impact of a magnetic field on magnetic correlations has also been studied in YBCO.  Measurements on detwinned crystals of $x=0.35$ and 0.45 showed an enhancement of the elastic incommensurate magnetic signal,\cite{haug09,haug10} with a greater relative increase for $x=0.45$ where the zero-field intensity is weaker.  A study by Stock {\it et al.}\cite{stoc09} on YBCO $x=0.33$ and 0.35 (twinned) crystals found no field enhancement of the elastic magnetic intensity; however, a  field was found to enhance the inelastic magnetic response for energies $\lesssim1$~meV in both samples.

\section{Impurity effects and related studies}
\label{sc:impurity}

\subsection{Contrasting Zn and Ni impurity effects on spin excitations}

Impurities in cuprates modify the carrier mobility and the spin correlations. Therefore, impurity substitution provides invaluable information on the interplay between superconductivity and magnetism in cuprates.  In particular, contrasting effects of nonmagnetic Zn and magnetic Ni impurity have been extensively studied.  From $\mu$SR studies by Nachumi {\it et al.},\cite{nach96}
a ``swiss cheese'' model was proposed for Zn-doped LSCO and YBCO: a Zn impurity locally destroys superconductivity and induces static spin correlation.  Furthermore, scanning-tunneling-spectroscopy studies on Zn-doped~\cite{pan00} and Ni-doped~\cite{huds01} Bi2212 have revealed that superconductivity is locally destroyed around a Zn atom, in contrast to weak degradation around a Ni atom.  As for the effect of impurities on magnetism, 
quasielastic magnetic peaks were first observed by Hirota {\it et al.}\cite{hiro98} 
in 1.2\%\ Zn-doped LSCO ($x=0.14$; $T_{c}=19$~K).  Subsequent neutron scattering studies by 
Kimura {\it et al.}~\cite{kimu03b} revealed that dilute Zn doping into optimally doped LSCO
induces excitations within the spin gap, much like the magnetic-field-induced in-gap excitations.\cite{lake01,chan09}  Based on the ``swiss-cheese'' model, it is likely that the in-gap excitations are produced locally by the Zn.  With increasing Zn concentration, the in-gap state is enhanced, and the spin correlations become more static as superconductivity is suppressed.

\begin{figure}[tb]
\begin{center}
\includegraphics[width=7.3cm]{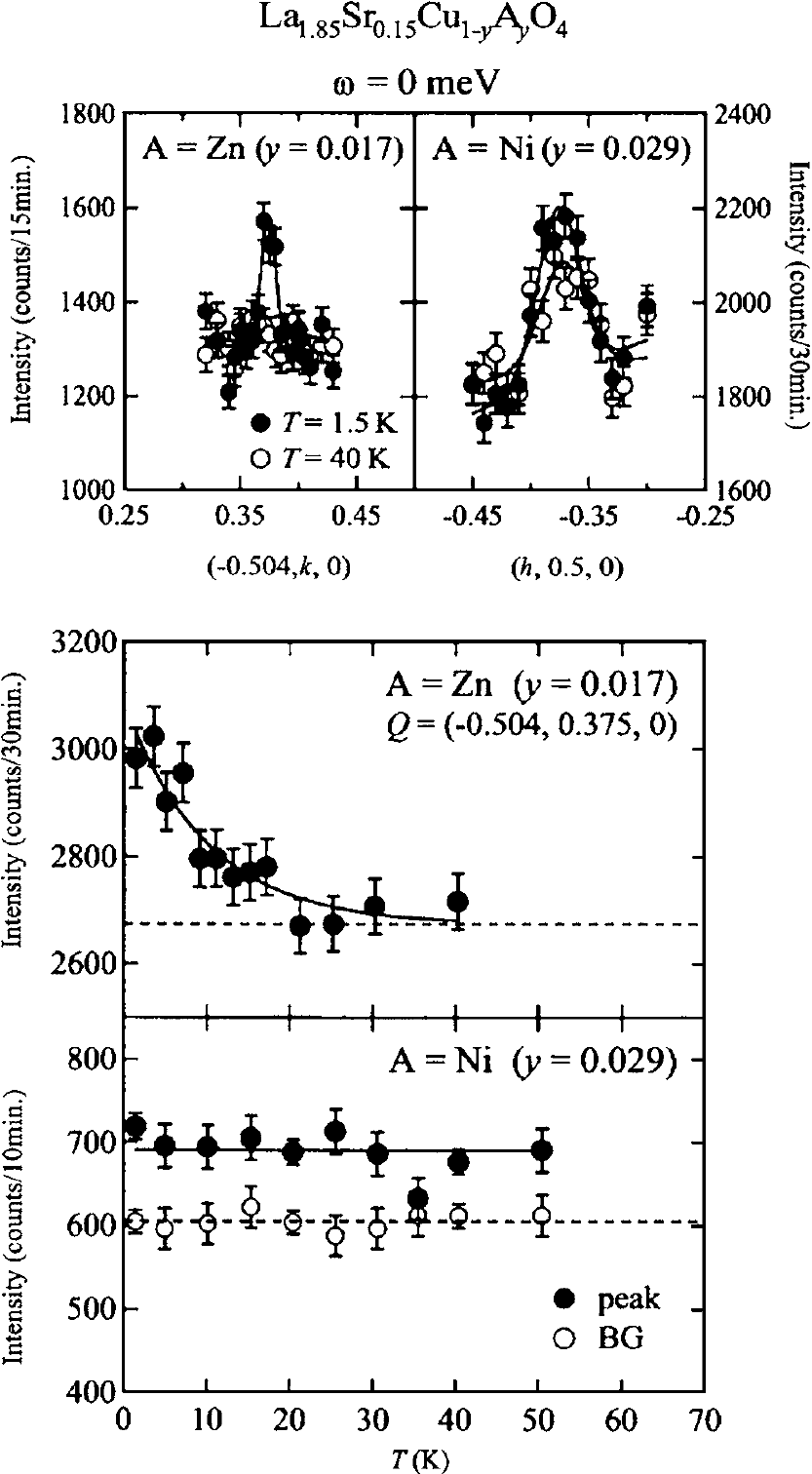}
\end{center}
\caption{(Top panels)
The incommensurate elastic peak profile
for La$_{1.85}$Sr$_{0.15}$Cu$_{1-y}$A$_{y}$O$_{4}$
with $A=$~Zn:$y=0.017$ ($T_{c}=16.0$~K) and $A=$~Ni:$y=0.029$ 
($T_{c}=11.6$~K).
(Bottom panels)
Temperature dependence of the intensity at peak position (closed circles)
and background (open circles).~\cite{kofu05}
}
\label{kofu_fig}
\end{figure}

The contrasting Zn and Ni impurity effects on spin correlations have been studied in optimally-doped LSCO by Kofu {\it et al.}\cite{kofu05} using neutron scattering.  Figure~\ref{kofu_fig} (top) shows peak profiles of the incommensurate elastic peak for La$_{1.85}$Sr$_{0.15}$Cu$_{1-y}A_{y}$O$_{4}$ with $A=$~Zn:$y=0.017$ ($T_{c}=16.0$~K) and $A=$~Ni:$y=0.029$ ($T_{c}=11.6$~K).  For Zn:$y=0.017$, a sharp elastic peak is observed at low temperatures, indicating that static, locally-AF, stripe-like order is induced by the Zn atoms.  In contrast, as shown in Fig.~\ref{kofu_fig} (bottom), a temperature-independent broad peak was observed for Ni:$y=0.029$.  The broad peak induced by Ni is attributed to inelastic signal detected at $\omega=0$ owing to coarse energy resolution.  Thus, dilute Ni impurities do not induce static spin order, whereas a smaller concentration of Zn impurities does.  In addition, Kofu {\it et al.}\cite{kofu05} found that dilute Ni doping does not induce the in-gap state, but, instead, reduces the spin-gap energy.  The reduction of the gap energy seems to correspond to the reduction of $T_{c}$ by Ni.  Such scaling behavior strongly suggests  the renormalization of the characteristic energy of the spin excitations.
A similar discussion was made by Tokunaga {\it et al.}\cite{toku97} based on NMR results.

\begin{figure}[tb]
\begin{center}
\includegraphics[width=7.3cm]{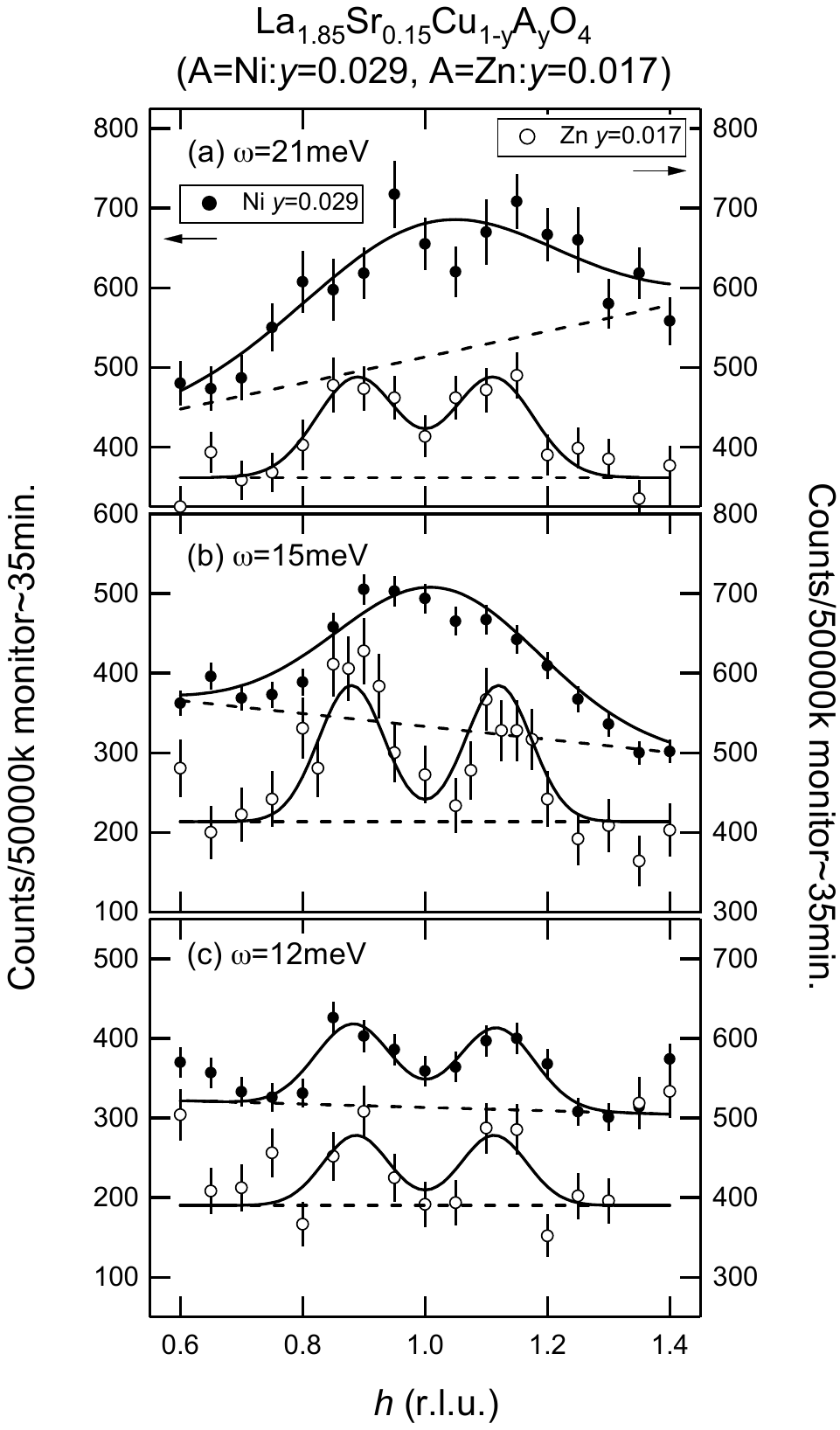}
\end{center}
\caption{(a)-(c) Profiles of constant-energy scan 
along the [100]$_{ortho}$ direction
around the AF zone center ($h=1.0$) at $T=11$~K
for La$_{1.85}$Sr$_{0.15}$Cu$_{1-y}$A$_{y}$O$_{4}$
with $A=$~Ni:$y=0.029$ (closed circles) 
and $A=$~Zn:$y=0.017$ (solid circles).~\cite{mats09}
The right axes for rhw Zn data are shifted 
for ease of comparison.
The dotted lines indicate background.
}
\label{matsuura_fig1}
\end{figure}

Matsuura {\it et al.}\cite{mats09} have performed extended neutron scattering studies on Zn- and Ni-doped LSCO to higher energies.  Figure~\ref{matsuura_fig1} shows the contrasting effect 
of the two types of impurities on the spin excitation spectra.  Zn doping does not drastically change the peak profile, which remains incommensurate up to at least 21~meV, while for Ni doping a broad commensurate peak appears by 15~meV.  The former effect is attributed to the local pinning effect of Zn, while latter is consistent with a reduction of the characteristic energy by Ni doping. 
Therefore, Matsuura {\it et al.}\cite{mats09} predicted a reduction of the crossover energy \ecross, at which a commensurate peak starts to appear.  They also proposed that Ni doping renormalizes the upper branch of the hour-glass dispersion, with the energy scale of the branch directly related to $T_{c}$.

\begin{figure}[tb]
 \begin{center}
   \includegraphics[width=7.3cm]{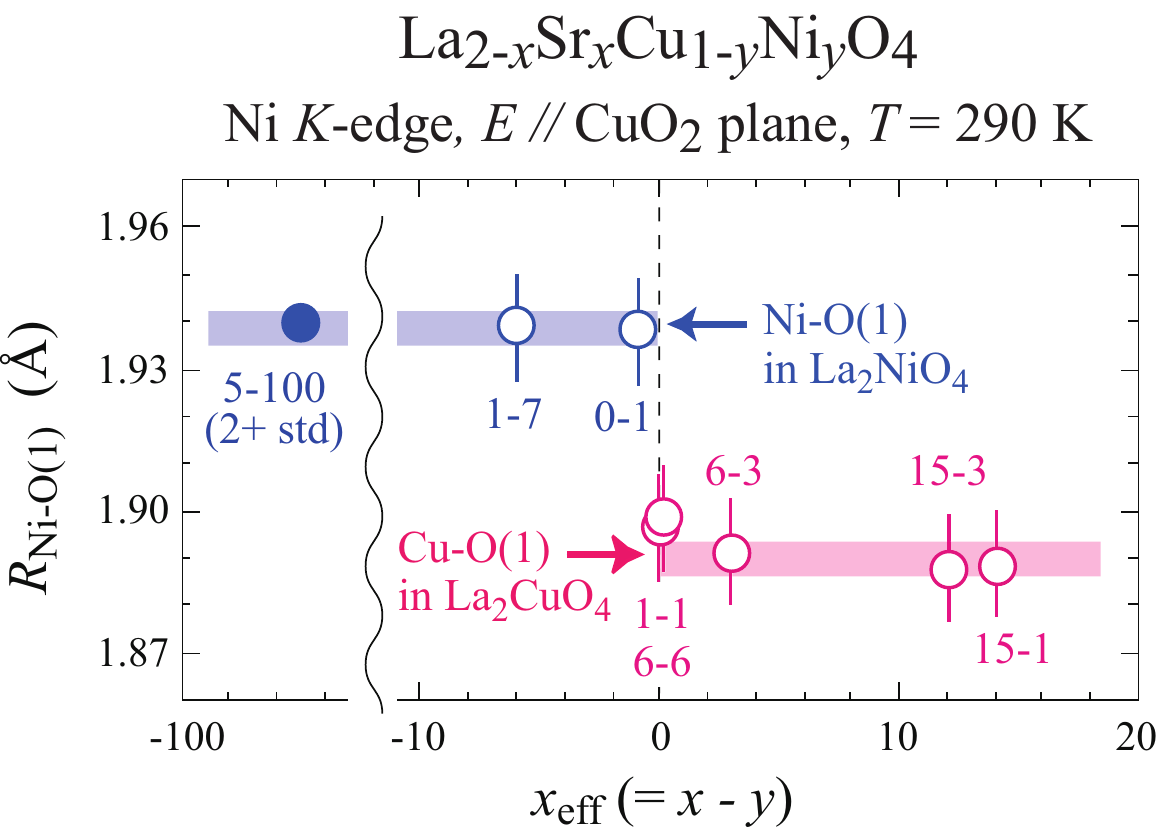}
    \caption{(Color online)
  The bond length $R_{\rm Ni-O(1)}$ in La$_{2-x}$Sr$_x$Cu$_{1-y}$Ni$_y$O$_4$ as a function of $x_{\rm eff}$ determined by Ni \textit{K}-edge XAFS measurements.~\cite{hira09} The hole and Ni concentrations are expressed as ($100x-100y$). The upper and lower arrows correspond to the Ni-O(1) distance in La$_2$NiO$_4$ and the Cu-O(1) distance in La$_2$CuO$_4$, respectively.
    }
    \label{hiraka-xafs}
  \end{center}
\end{figure}

What is the origin of the different impacts on high-$T_{\rm c}$ superconductivity and spin dynamics between Zn and Ni impurities?  Is it simply associated with the non-magnetic nature of Zn$^{2+}$ ($S=0$) and the magnetic nature of Ni$^{2+}$ ($S=1$)?  One possibility is that each Ni dopant might tend to localize a hole near it.\cite{bhat92,mend94,naka98c,wata03,hira05,mats06,hira07}  To test this idea, polarized X-ray-absorption-fine-structure (XAFS) measurements were performed 
at the Ni $K$-edge on single crystals La$_{2-x}$Sr$_{x}$Cu$_{1-y}$Ni$_{y}$O$_4$.\cite{hira08,hira09}  The measurements revealed two distinct types of Ni-dopant site, with signatures given by shifts in the edge energy and in $R_{\rm Ni-O(1)}$, where $R_{\rm Ni-O(1)}$ is the 
interatomic distance between Ni and in-plane oxygen O(1) (Fig.~\ref{hiraka-xafs}).
A state with a higher valence than Ni$^{2+}$ was observed for $x_{\rm eff} (\equiv x-y)\ge 0$, 
while a state consistent with Ni$^{2+}$ was found for $x_{\rm eff}<0$.  The higher-valence state is most likely described by Ni$^{2+}\underline{L}$ with $S_{\rm eff}=1/2$ ($\underline{L}$ represents a ligand hole), rather than Ni$^{3+}$, as proposed previously.  In other words, a hole is strongly bound around Ni on neighboring oxygen orbitals, thus forming a Zhang-Rice doublet state; this picture is supported by theory.\cite{tsut09}  It is also consistent with recent experimental work by Tanabe \textit{et al.}\cite{tana10} based on specific heat and $\mu$SR measurements.  They claimed that a Ni dopant changes its character from a strong hole absorber, in the underdoped region, to a Kondo scatterer in the overdoped, metallic region.  Therefore, the impact of a hole-trapped Ni impurity on the Cu-spin network must be small in magnitude and extended in space, resulting in the renormalization of the energy scale for magnetic fluctuations.\cite{kofu05,mats09}
This is a possible reason for the smaller effect of hole-trapped Ni on high-$T_{c}$ superconductivity, as compared to nonmagnetic Zn.

\subsection{Stability of stripe order}

At ambient pressure, the lattice distortion associated with the CDW order has so far been observed only in the LTT phase.  (We note that CDW order was observed in the pressure-induced HTT phase of LBCO by X-ray diffraction measurement.\cite{huck10})   In particular, there has been no diffraction evidence reported for CDW order in the LTO phase,\cite{fuji02,kimu04} though we have seen that there is evidence for SDW order.  Thus, the degree to which CDW and SDW order appear together is in question.  To shed more light on this issue, the impurity effect on both spin and charge orders has been investigated in Fe-doped LSCO (Fe-LSCO),\cite{fuji08b,fuji09a,fuji09b} where the average LTO crystal structure is not affected by the small concentration of Fe dopants. 

\begin{figure}[t]
\begin{center}
\includegraphics[width=8.2cm]{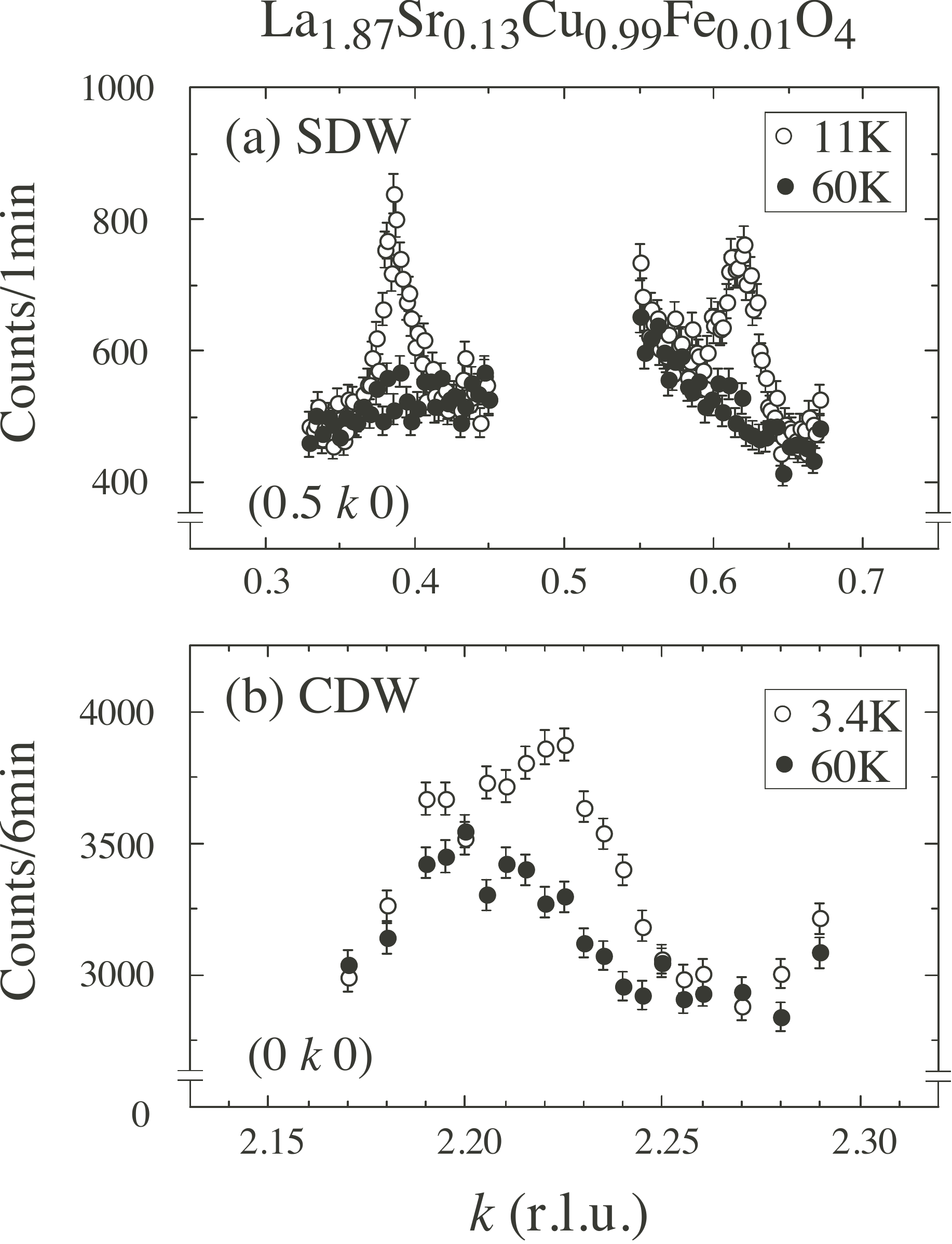}
\end{center}
\caption{Incommensurate peaks from (a) SDW and (b) CDW orders in the LTO phase of La$_{1.87}$Sr$_{0.13}$Cu$_{0.99}$Fe$_{0.01}$O$_4$ measured at low-temperature (closed circles) and high- temperature (open circles).\cite{fuji08b}}
\label{fg:fe}
\end{figure}

Peak profiles for SDW order in La$_{1.87}$Sr$_{0.13}$Cu$_{0.99}$Fe$_{0.01}$O$_4$ are shown in Fig.~\ref{fg:fe}(a). Four incommensurate (IC) peaks are located at $(0.5, 0.5\pm\delta, 0)$, $(0.5\pm\delta, 0.5, 0)$ positions with $\delta=0.115\pm0.003$, consistent with results for Fe-free LSCO;  however, the volume-corrected intensity is much stronger in the La$_{1.87}$Sr$_{0.13}$Cu$_{0.99}$Fe$_{0.01}$O$_4$ system.\cite{fuji08b} Indeed, the SDW order in Fe-free LSCO was difficult to detect under the identical experimental setups. The onset temperature for the appearance of SDW peaks ($T_m$) of $\sim50$~K is slightly higher than $T_m\sim40$~K in LSCO. The enhancement of peak intensity and the ordering temperature suggest the stabilization of SDW order by Fe-doping. 

Even more importantly, an IC peak from CDW order was detected in the Fe-doped LSCO sample, as shown in Fig.~\ref{fg:fe}(b). At low temperature, a clear enhancement of intensity was observed at $(0, 2\pm\epsilon, 0)$ with $\epsilon=0.224\pm0.002\approx2\delta$. Since the well-defined CDW order has not been detected in the Fe-free LSCO, the observation of CDW order in the present system indicates the inducement (or strong enhancement) of CDW order by Fe-doping. Surprisingly, the onset temperature for the appearance of CDW order is close to that in the LBCO system, in which the well-stabilized CDW order is realized at low temperature.\cite{fuji04} Furthermore, the volume-corrected intensity in the present Fe-doped LSCO is half of that in LBCO. Therefore, the Fe doping has stabilized bulk CDW order, not just local patches. Indeed, the coherence length for the  lattice distortion evaluated from the width of the CDW peak,  60 \AA, exceeds the mean distance between nearest-neighbor Fe ions ($\sim38$~\AA).\cite{fuji09a} Thus, bulk stripe order, which is identical to that observed in the LTT phase, indeed exists in the present system with LTO structure, and the static stripe order can be realized by impurity substitution as is expected from the the stripe-pinning picture. 

\subsection{Spin-impurities in the overdoped region}

\begin{figure}
\begin{center}
\includegraphics[width=8.2cm]{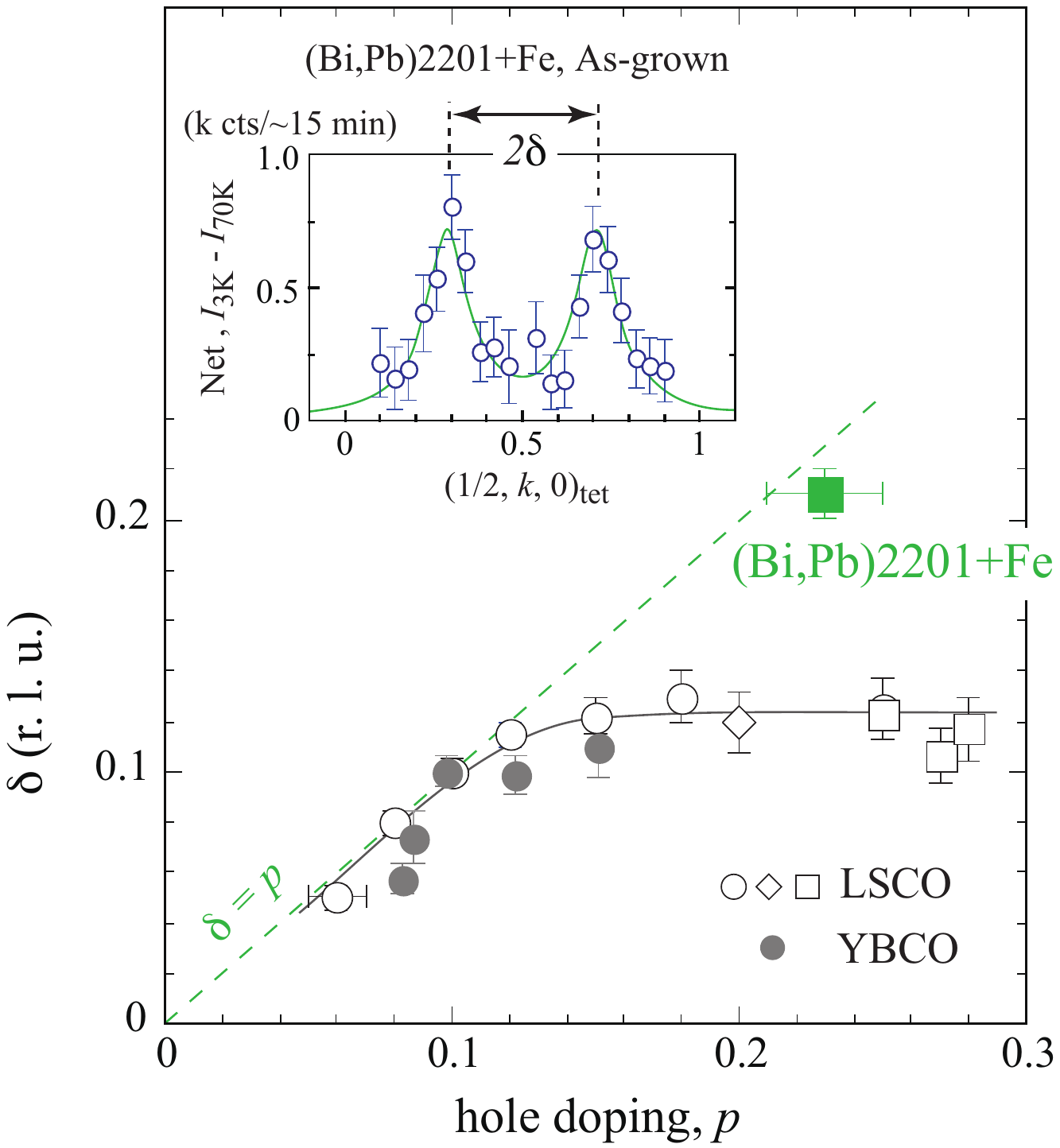}
\end{center}
\caption{(Color online)
Comparison of $\delta$ of the parallel spin modulation between Fe-doped (Bi,Pb)2201, pristine LSCO, and YBCO.~\cite{hira10}
The inset shows magnetic, incommensurate elastic peaks from Bi$_{1.75}$Pb$_{0.35}$Sr$_{1.90}$Cu$_{0.91}$Fe$_{0.09}$O$_{6+z}$ (as-grown, overdoped sample). 
Dynamical data are plotted for LSCO ($\omega=3-6$ meV) and YBCO ($\omega\ll$ (resonance energy)).
The solid line is a guide for LSCO and YBCO data.  The dashed line shows the linear relation $\delta = x$.
}
\label{fg:waki2}
\end{figure}

The Fe-doping in the underdoped region stabilizes stripe order, as presented above.  In contrast, Fe-doping effects in the overdoped regime have suggest a different behavior.  Hiraka {\it et al.}\cite{hira10} substituted Cu sites with Fe spins in overdoped Bi$_{1.75}$Pb$_{0.35}$Sr$_{1.90}$CuO$_{6+z}$, a system for which both neutron scattering and $\mu$SR experiments have never detected any sign of magnetic correlations.  The Fe substitution resulted in short-ranged incommensurate static spin correlations.  Interestingly, the observed incommensurability of $\delta = 0.21$ is far beyond the upper limit of $\delta \sim 0.13$ observed so far for LSCO and YBCO.  Instead, $\delta$ is close to the hole concentration $p \sim 0.23$ estimated from ARPES experiments.  Figure~\ref{fg:waki2} shows the comparison of $\delta$ in various compounds.  Wakimoto {\it et al.}\cite{waki10} concluded from the magnetic field dependences of resistivity and spin correlations in the Fe-doped Bi system that RKKY coupling between the Fe-spins via conduction electrons is the most plausible origin of the incommensurate spin correlation, which is rather different from the underdoped region. Substitution of Cu sites with Fe spins for overdoped LSCO induces a similar incommensurate static spin correlation. In this case, the $\delta$ is also larger than 0.12 and monotonically increases with increase of carrier concentration up to near the upper boundary of superconductivity.\cite{he11} ARPES experiments on the same system predict a similar doping dependence of the $\delta$ in the overdoped region based on Fermi-surface nesting.

These results strongly suggest a change in the degree of electronic correlation between the underdoped and overdoped regions. Concerning the orbital character of carriers in cuprates, comprehensive studies have concluded that the doped holes predominantly enter into the oxygen $2p$ orbital at least up to optimal doping.\cite{nuck89,bian87,chen91,zhan88,peet09}  As a result, the unusual physical properties of underdoped cuprates have been analyzed mainly by ascribing a single orbital character to the doped holes. However, in the overdoped cuprates the orbital character is not fully understood, even though distinct doping dependencies of x-ray absorption\cite{chen92,peet09} and optical reflectivity spectra\cite{uchi91} between the underdoped and overdoped regions suggest a change in the oxygen $2p$ orbital character with overdoping.

In a recent study on single crystals of LSCO covering a broad range of doping, high-resolution Compton scattering measurements  confirmed the change of orbital state with overdoping.\cite{saku11} The holes in the underdoped regime are found to primarily populate the O $2p_{x}/p_{y}$ orbitals.  In sharp contrast, holes mostly enter Cu-$e_{2g}$ orbitals in the overdoped system.
These studies in the overdoped region reveal how the standard Zhang-Rice picture of doped holes in this strongly correlated cuprate system evolves into a more conventional mean-field description of electronic states as correlations weaken with doping.

\section{Electron doping}
\label{sc:edope}

\subsection{Low-energy excitations}

In electron-doped ($n$-type) high-$T_c$ cuprates there remain more unsolved issues than in hole-doped ($p$-type) cuprates.\cite{armi10} Here, we briefly review doping dependence of low-energy magnetic excitations of $n$-type cuprates and compare the results with those for $p$-type cuprates. Low-energy magnetic excitations exhibit commensurate peaks centered at \qaf\ for both antiferromagnetic and superconducting phases, in contrast to the incommensurate ones for the $p$-type superconducting cuprates.\cite{yama03, fuji03}  Carrier doping or annealing under reduced atmosphere broadens the peak width of the commensurate magnetic signal. 

Wilson {\it et al.}\cite{wils06b} performed inelastic neutron scattering experiments on Pr$_{1-x}$LaCe$_{x}$CuO$_{4+\delta}$ (PLCCO) with $x=0.12$ and several different oxygen concentrations at various temperatures ($T$) with energies up to $\sim4$ meV. They found a crossover of dimensionality in the spin fluctuations from three dimensional in the AF phase to two dimensional in the SC phase. They also found a $T$-insensitive magnetic excitation spectrum near the optimally doped SC phase. 

An alternative approach was taken by Motoyama {\it et al.}\cite{moto07} in a study of \ncco\ (NCCO) where they performed neutron total scattering measurements.  From the thermal evolution of instantaneous spin-spin correlation length, they extracted the effective spin stiffness.  The spin stiffness is well-defined in the AF phase, decreasing with increasing $x$, and eventually reaching zero at the AF-SC boundary. Within the SC phase, the spin correlation length is temperature independent, with a magnitude comparable to the superconducting coherence length.

\begin{figure}[t]
\begin{center}
\includegraphics[width=7.7cm]{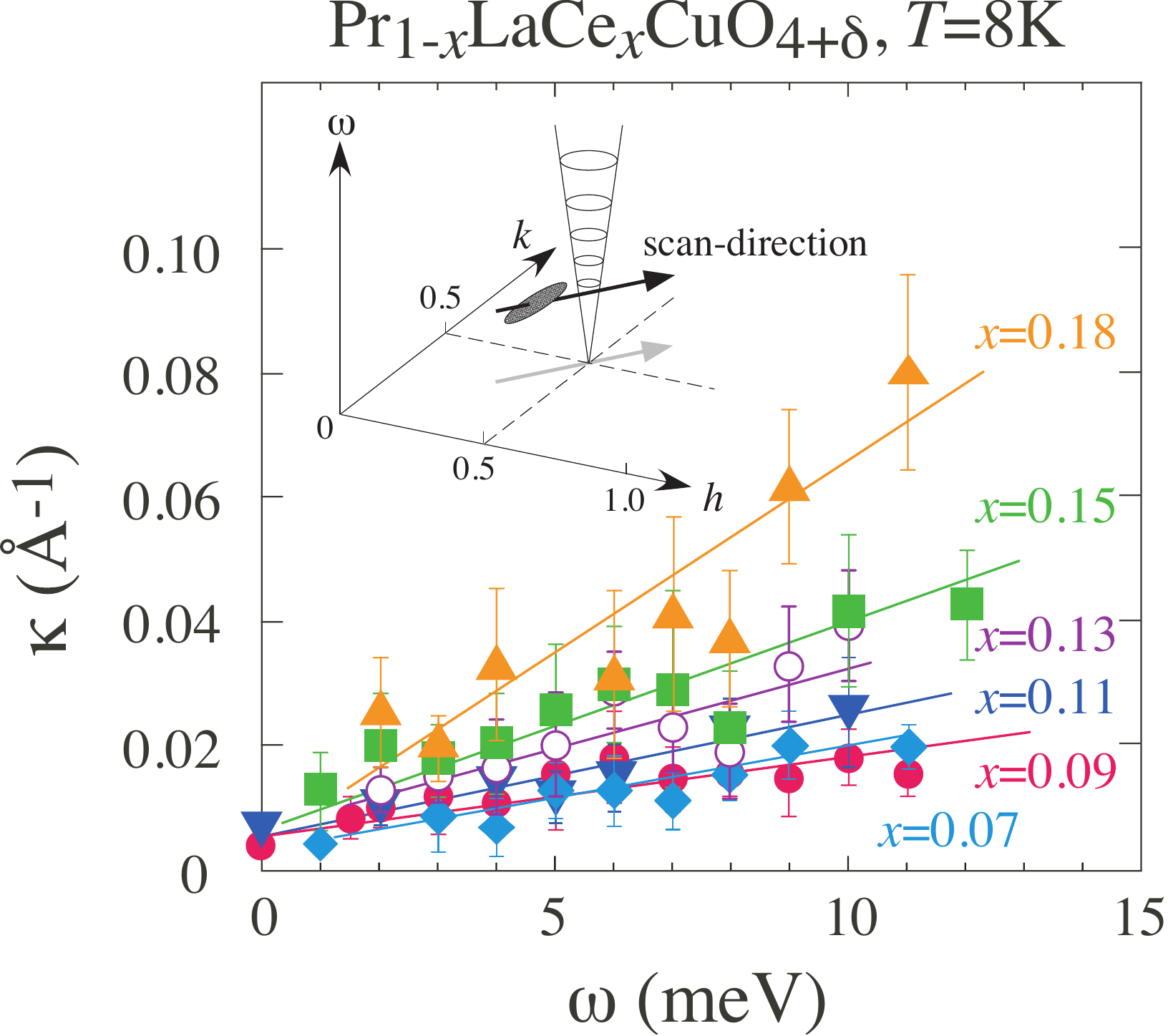}
\end{center}
\caption{(Color online) 
$\omega$-dependence of resolution corrected peak-width (half width at half maximum) $\kappa$ of commensurate peak for Pr$_{1-x}$LaCe$_{x}$CuO$_{4}$ with $x=0.07$, 0.09, 0.11, 0.15 and 0.18, from Ref.\cite{fuji08}.}
\label{fg:ed1}
\end{figure}

\begin{figure}[t]
\begin{center}
\includegraphics[width=7.7cm]{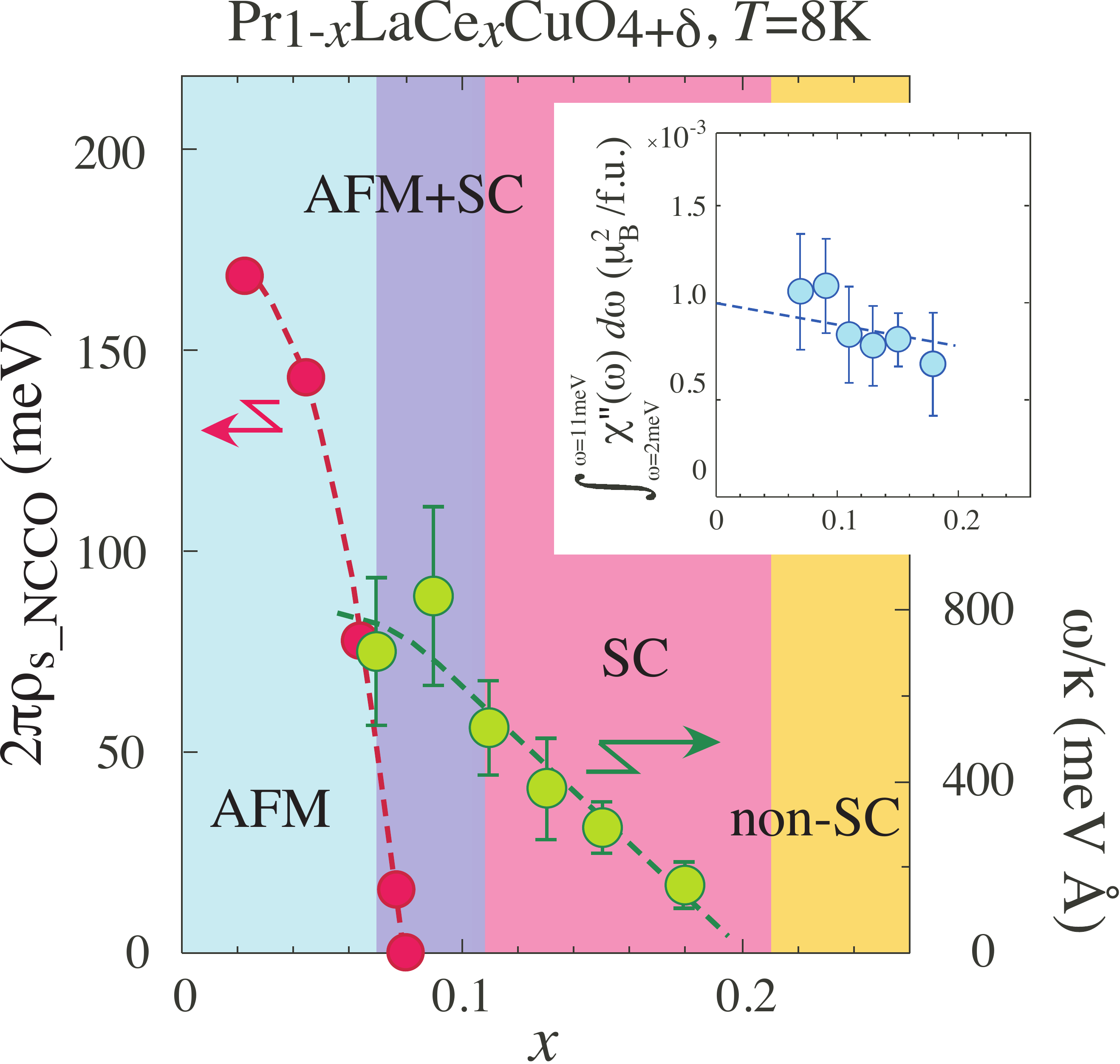}
\end{center}
\caption{(Color online)  Doping dependence of the low energy spin fluctuations: (a) the spin stiffness, $\omega/\kappa$, and (b) the partial spectral weight obtained by integrating $\chi{\prime\prime} (\omega)$ from 2 meV to 11 meV, as a function of $x$. Dashed lines are guides to the eye. From Ref.\cite{fuji08}}
\label{fg:ed2}
\end{figure}

Fujita {\it et al.}\cite{fuji08} performed comprehensive neutron inelastic scattering experiments on PLCCO over a wide doping region, extending close to the upper critical concentration for superconductivity, to elucidate the nature of low-energy spin fluctuations in the SC phase.  Looking at how the effective dispersion, the excitations appear to form a filled cone with its tip at \qaf\ and $\hbar\omega=0$.  From constant energy scans, one can extract the half-width-at-half-maximum as a function of momentum, $\kappa$, which is plotted in Figure~\ref{fg:ed1} for a series of samples.  For each concentration, $\kappa$ increases linearly with $\omega$ up to $\sim12$~meV. The inverse of the slope, $\rho_{\rm s} = \omega/\kappa$, defines the low energy spin stiffness, which decreases linearly with increasing $x$ within the entire SC region, as shown by the green circles in Fig.~\ref{fg:ed2}. Extrapolation indicates that $\rho_{\rm s}$ goes to zero near the SC/non-SC phase boundary ($x_{\rm c} \sim 0.21$). It is to be noted that the critical value of $x_{\rm c}\sim 0.21$ is well below the percolation limit, $\sim 0.41$,\cite{mang04a, newm00} in the two-dimensional square-lattice spin system. Therefore, the observed degradation is not explained by the simple model of randomly-diluted quantum spins.

Such contrasting doping dependences of spin stiffness defined either by the thermal evolution of the instantaneous (energy-integrated) spin correlation length or by the energy dependence of the low-energy spin dispersion width strongly suggests the contrasting nature of spin fluctuations between the AF and SC phases, which is possibly attributed to the existence of a quantum critical point (QCP) at the phase boundary in the electron-doped cuprates. Furthermore, Fujita {\it et al.}\cite{fuji08} found a linear relation between $T_{\rm c}$ and the characteristic energy $\Gamma$, at which the $q$-integrated intensity around \qaf\ shows a maximum, consistent with the doping dependence of the low-energy spin stiffness mentioned above.

In the case of $n$-type PLCCO, the overall spectral weight does not change much with doping concentration, even in the overdoped region (see the inset of Fig. ~\ref{fg:ed2}). This suggests that even in the SC phase localized spin character remains in the low energy region.  This is one possible reason why a simple band model cannot reproduce the commensurate nature of the spin fluctuations in the $n$-type cuprate.\cite{krug07} The continuous degradation of the effective magnetic interactions and the peak-broadening upon doping reflects a more homogeneous electronic state in the electron-doped cuprates than in the hole-doped ones. Such contrasting behavior can be understood if we consider the difference in the orbital character of doped carriers in the $n$-type and the $p$-type cuprate. In the case of $n$-type, the doped electrons go into the Cu 3$d$-orbitals\cite{tran89b} and continuously degrade the spin correlations as the excess electron concentration increases. On the other hand, in the case of $p$-type, doped holes first enter into O 2$p$-orbitals. Upon overdoping, however, the holes start entering also into Cu 3$d$-orbitals yielding two types of locations for holes, that may cause the inhomogeneous phase separation.\cite{yama98a,waki07b,uemu03,tana05}. 

The resonance like enhancement of magnetic signal in the low-energy spin excitation has been independently studied by two groups. For PLCCO with $T_c=24$~K, Wilson {\it et al.}\cite{wils06} found a resonance-like peak at \qaf\ with the energy $E_r \sim 11$~meV$\sim5.3k_{\rm B}T_c$, and they claimed the similar relation between $E_r$ and $T_c$ for both $p$-type and $n$-type cuprates. On the other hand, for NCCO with $T_c=25$~K, Yu {\it et al.}\cite{yu10b} revealed two distinct magnetic energy scales in the superconducting state: 6.4 meV and 4.5 meV, both of which are much smaller than the resonance energy for PLCCO. According to their interpretation, the former energy is the maximum superconducting gap, but the origin of the latter has remained unexplained. They also discussed that the latter energy is consistent with a resonance and with the recently established universal ratio of resonance energy to superconducting gap in unconventional superconductors. However, Zhao {\it et al.}\cite{zhao07b} independently performed a neutron inelastic scattering experiment on NCCO with $T_c=25$~K and found a resonance-like enhancement of magnetic signal at $\sim9$~meV. Therefore, the final conclusion on this issue is controversial, possibly due to the experimental challenge to account for the effects of crystal field excitations from the rare-earth ions.  Related with such contrasting properties between PLCCO and NCCO, it should be noted that in PLCCO no clear spin gap has been observed, while a finite energy gap was confirmed to open below $T_c$ in the NCCO system.\cite{yama03} 

Very recently, Zhao {\it et al.}\cite{zhao11} have combined neutron scattering and scanning tunneling spectroscopy (STS) measurements on a pair of PLCCO samples.  They found evidence both for a resonance peak and in-gap excitations.  The STS results indicate that the spin resonance (in-gap signal) is correlated (anti-correlated) with the magnitude of the superconducting gap, which varies on the scale of nanometers.

\subsection{High-energy excitations}

As shown in \S\ref{sc:magex}, neutron-scattering experiments have revealed a remarkable similarity of the overall magnetic excitation spectrum in the hole-doped cuprates.  In the electron-doped superconducting cuprates however, only two independent neutron scattering experiments have explored the overall spin excitation spectrum. The experiment on PLCCO with $x=0.12$ sample ($T_c=21$~K) by Wilson {\it et al.}\cite{wils06c} found that the effect of electron-doping is to cause a wave-vector broadening in the low-energy ($E<80$ meV) commensurate spin fluctuations at \qaf\ and to suppress the intensity of spin-wave-like excitations at high energies ($E>100$ meV). The obtained magnetic dispersion is anomalous. If they fit it by a two dimensional spin wave dispersion, the nearest neighbor interaction $J$ is obtained to be $\sim162$~meV which is much larger than that of non-doped Pr$_2$CuO$_4$  ($J\sim121$ meV)~\cite{bour97c}. Furthermore, the local spin susceptibility $\chi''(\omega)$ is anomalously smaller than those of non-doped La$_2$CuO$_4$ and hole-doped La$_{1.875}$Ba$_{0.125}$CuO$_4$.

\begin{figure}[t]
\begin{center}
\includegraphics[width=7.7cm]{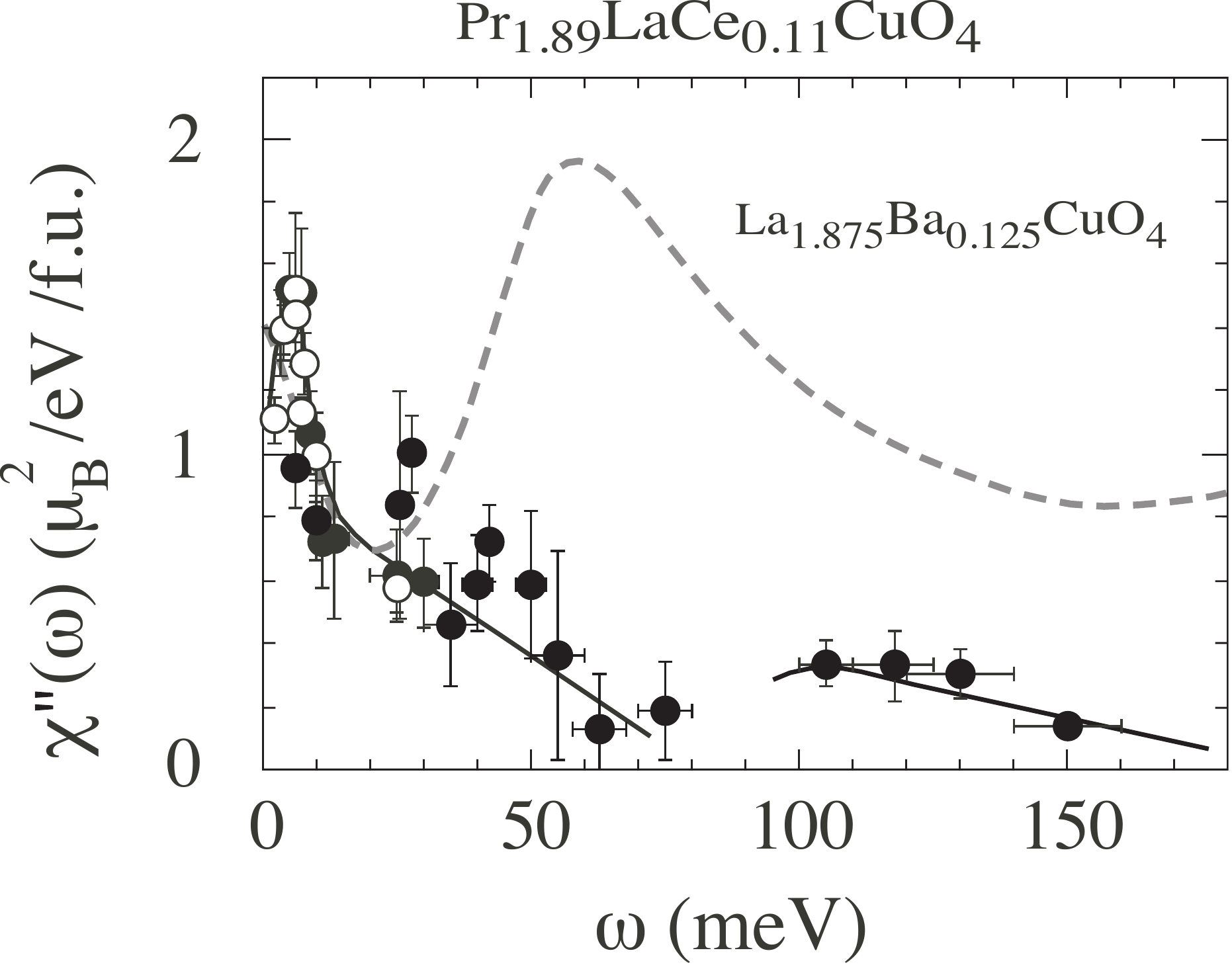}
\end{center}
\caption{Energy dependence of local susceptibility $\chi$$^{\prime}$$^{\prime}$($\omega$) for Pr$_{0.89}$LaCe$_{0.11}$CuO$_{4}$, from Ref.\cite{fuji06b}. Results for La$_{1.875}$Ba$_{0.125}$CuO$_4$ is referred by a dashed line. }
\label{fg:ed3}
\end{figure}

\begin{figure}[b]
\begin{center}
\includegraphics[width=7.7cm]{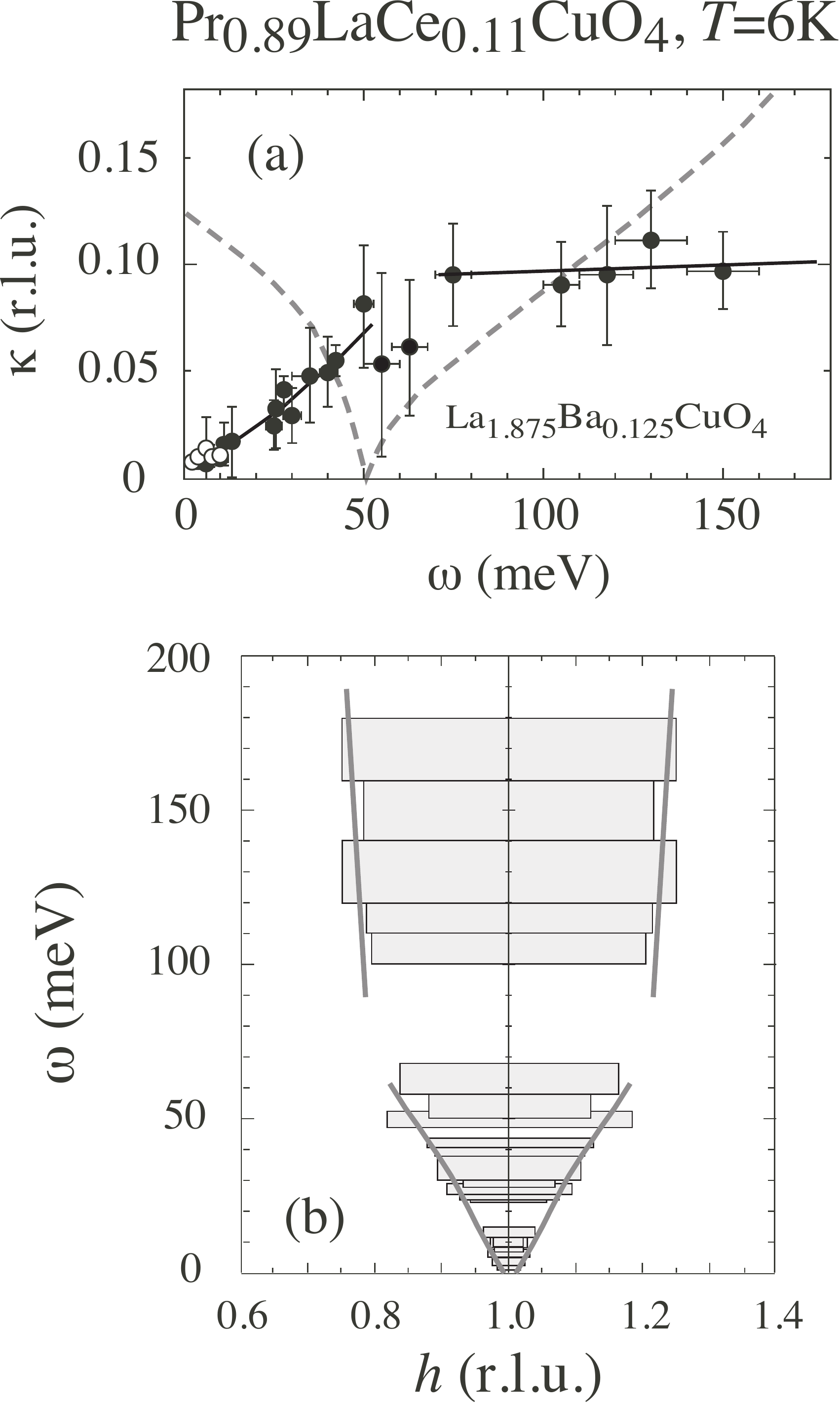}
\end{center}
\caption{(a) Energy dependence of the resolution corrected peak width $\kappa$ for Pr$_{0.89}$LaCe$_{0.11}$CuO$_{4}$, from Ref.\cite{fuji06b}, and (b) the overall shape of spin excitation spectrum  In the lower figure, the horizontal(vertical) length of the rectangle represents a fitted result of full-width at half-width with assuming a single Gaussian  function (sliced energy range for the analysis). }
\label{fg:ed4}
\end{figure}

An independent high-energy inelastic neutron scattering experiment was performed by Fujita {\it et al.}\cite{fuji06b} using a large number of single crystal of optimally doped PLCCO ($x=0.12$, $T_c=25.5$~K). Magnetic intensities were again confirmed to persist around \qaf\ in a wide energy range up to $\sim180$~meV, except for a gap at $\hbar\omega\sim60$~meV, as shown in Fig.~\ref{fg:ed3}.  As the energy transfer $\hbar\omega$ increases, the commensurate peak broadens and weaken in intensity, consistent with the results of Wilson {\it et al.}\cite{wils06c}  However, the observed high-energy excitations, at least up to 180~meV, are difficult to understand by the conventional spin-wave approximation because the expected upper bound energy from the value of $J$ evaluated in the low-energy region is only $\sim120$~meV. Therefore, the authors predict a different nature for the high-energy spin excitations from that of the low energy excitations in the electron-doped cuprate.  In fact, in the high-energy region between 100 and 180~meV, the {\bf q} width shows little variation with $\hbar\omega$ and is comparable to the value at $\sim60$~meV; the overall {\bf q}-dependence of the spin excitations is approximately pencil-shaped, as illustrated in Fig.~\ref{fg:ed4}.  The spin excitations extending up to the high-energy region remind us of  similar spectra observed in the nearly antiferromagnetic metals Cr$_{0.95}$V$_{0.05}$~\cite{hayd00} and Mn$_{2.8}$Fe$_{0.2}$Si~\cite{tomi87}. 

The persistence of the high-energy spin fluctuations around \qaf\ is consistent with a result from Fermi liquid theory\cite{yana01} which shows that the spin fluctuations in the narrow range of momentum space around \qaf\ weakens the pairing interaction, so that $T_c$ becomes lower compared to the case of the hole-doped system.  Combined with the fact that commensurate low-energy spin fluctuations can not  be reproduced by a band model, this similarity suggests that the itinerant nature of electrons is the possible origin of high-energy spin excitations. The dual structure of the spin excitations would reflect a crossover in the nature between the itinerancy and the localization of electrons, namely, the high-energy part of the excitations is a response of quasiparticles, while the low-energy part involves localized spins. This energy-dependent feature from the two spin degrees of freedom is different from what has been discussed in the hole-doped system as introduced in \S\ref{sc:magex}. (In the hole-doped superconducting system, it has been discussed that high-energy dispersive magnon-like modes are a sort of remnant excitation of the AF phase, while the low-energy spin dynamics, including the resonance feature, might originate from the response of quasiparticles. Indeed, phenomenological theory, which treats both itinerant fermions and local spins have well
reproduced the overall spin susceptibility in YBCO.\cite{bang09})  Therefore, even though both systems show evidence of dual character in the excitation spectrum and the energy for the separation is comparable, the differences in the dispersion of spin excitations between the $n$- and $p$-type systems suggest the different origins of the dual nature.  

\section{Other topics}
\label{sc:other}

\subsection{Exotic magnetic order in pseudo-gapped states}

Varma has proposed\cite{varm97,varm06} that valence fluctuations between a Cu atom and its O neighbors should lead to complicated patterns of current loops.  These current loops should generate magnetic moments that break time-reversal symmetry and four-fold rotational symmetry, but that preserve translational symmetry.  Because of the translation symmetry, magnetic scattering from ordered loop currents should occur only at reciprocal lattice vectors.  The form factor is {\bf Q} dependent and falls off rapidly with $Q$ because of the spatially extended nature of the magnetization density.

Motivated by Varma's predictions, Bourges and coworkers\cite{fauq06,mook08,bale11} have performed a series of polarized neutron diffraction experiments to test for unusual magnetic order.  These are very challenging experiments, as on must detect a small magnetic signal on top of a substantial diffraction intensity from the chemical order.  Initial measurements on several good quality, underdoped YBCO crystals revealed a small enhancement of the spin-flip cross section relative to the non-spin-flip cross section at temperatures comparable to the pseudogap regime ($T_{\rm mag}\sim200$~K for YBCO $x\sim0.6$).\cite{fauq06}  Based on the original current loop model, one would expect the magnetic moments to point along the $c$ axis; however, the experiment found that the effective moment direction was approximately $45^\circ$ away from $c$.  A collaborative experiment with Mook\cite{mook08} on a large YBCO $x=0.6$ crystal essentially confirmed these observations.  For measurements at a given reciprocal lattice vector, the identified spin-flip signal has the same {\bf q} width as the nuclear scattering, including along the $c$ axis, implying long-range order.

A recent study on YBCO $x=0.45$ and 2\%\ Zn-doped $x=0.6$ crystals, where static or quasi-static incommensurate spin order was previously observed at low temperature,\cite{hink08,such10} found a reduced magnitude of the spin-flip signal at Bragg wave vectors.  For the $x=0.45$ crystal, the onset temperature was also reduced.

A study of LSCO $x=0.085$ identified a spin-flip signal at the tetragonal (100) reflection; however, in contrast to YBCO, the signal was independent of $Q_z$ (indicating 2D character) and had a correlations length of $\sim10\ \AA$ within the CuO$_2$ planes.\cite{bale10}  The onset temperature was 120~K.

Greven's group\cite{li08,li10b,yu10} has studied unique crystals of HgBa$_2$CuO$_{4+\delta}$.  Li {\it et al.}\cite{li08} identified a spin-flip signal, similar to that in YBCO,\cite{fauq06} for three underdoped compositions.  More recently, intriguing inelastic responses have been reported.  After first reporting an antiferromagnetic resonance at 56 meV in a sample with $T_c=96$~K (measured with unpolarized neutrons),\cite{yu10} polarized neutron scattering has been used to identify a weakly dispersing branch that connects with the resonance and has roughly constant intensity across the Brillouin zone.\cite{li10b}  He and Varma\cite{he11} argue that this branch is a collective mode of the loop-current state.  While this new feature is interesting, it is important to note that studies on other cuprates have found no evidence for such a weakly dispersing magnetic mode.  In particular, a recent polarized-beam inelastic study\cite{head11} of YBCO $x=0.9$ did find magnetic scattering close to the antiferromagnetic wave vector in the normal state, spread over the energy range of 10--60 meV; however, the magnetic response did not extend outside of the AF Brillouin zone.

\subsection{Exploring new systems}
Neutron-scattering studies of high-$T_c$ superconductors reveal a close correlation between local antiferromagnetism and the superconductivity.  For instance, the hour-glass-shaped dispersion commonly observed in the superconducting phase of LSCO, YBCO and Bi2212 systems suggests the existence of a universal nature to the spin correlations in hole-doped cuprates.\cite{tran04,hayd04,fauq07}  However, it would only take one counter example to disprove the trend. We have hence started a systematic study of the spin excitations in the single-layer Bi$_{2+x}$Sr$_{2-x}$CuO$_{6+\delta}$ (Bi2201) system, in which the carrier concentration can be controlled by substituting Bi ions onto the Sr site. 

\begin{figure}[t]
\begin{center}
\includegraphics[width=7.7cm]{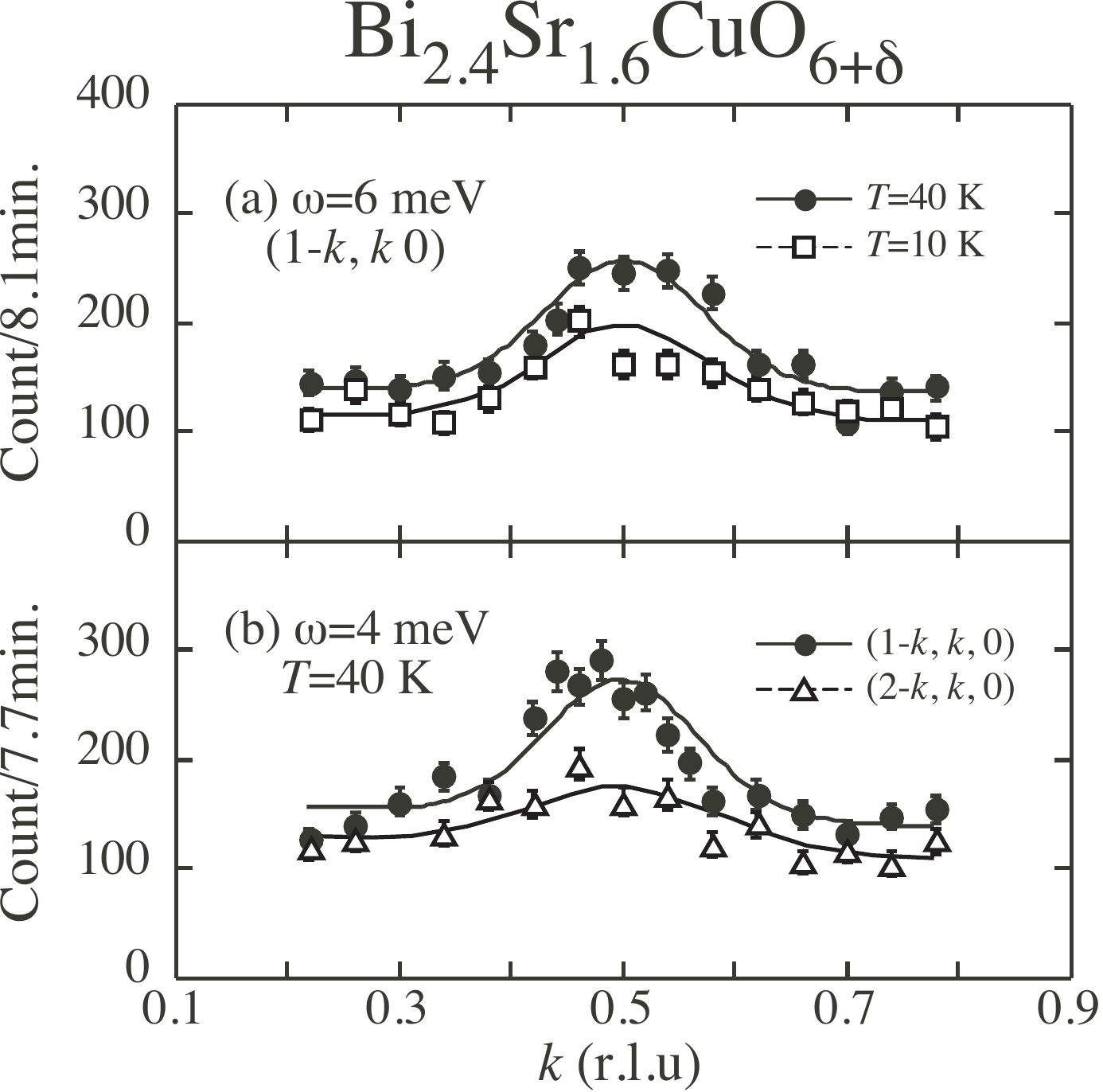}
\end{center}
\caption{\label{raw-data}Spin excitation spectra\cite{enok10} for Bi$_{2.4}$Sr$_{1.6}$CuO$_{6+\delta}$ for (a)$\hbar\omega=4$ meV and (b) 6 meV. Filled-circles and open-squares are the data measured around $(0.5,0.5,0)$ at $T=40$ K and 10 K, respectively. Open-triangles are the data measured around $(1.5,0.5,0)$ at $T=40$ K.}
\label{fg:bi2201}
\end{figure}

Figure~\ref{fg:bi2201} shows the inelastic neutron-scattering profile measured at $\hbar\omega=4$ and 6 meV on Bi$_{2.4}$Sr$_{1.6}$CuO$_{6+\delta}$, which is a lightly-doped sample.\cite{enok10}  In each measurement, the scan was made along along the $[1,-1,0]$ through \qaf.  A well-defined single peak was observed at $T=40$ K (closed-circled), and a similar result was obtained from a scan   along the $[1,0,0]$.  In order to clarify the origin of the signal, we next examined the temperature dependence of the signal and measured the profile around other magnetic zone centers with larger \textbf{Q}.  In Fig.~\ref{fg:bi2201}(a), the inelastic spectrum measured at $T=10$~K and $\hbar\omega=6$~meV is shown by open squares.  The peak intensity at 10 K is weaker than that at 40K, showing clearly the thermal evolution of the signal.  This result suggests that the signal originates from an intrinsic excitation, such as a magnon or phonon.  The comparison between the spectrum measured around ${\bf Q}=(0.5, 0.5)$ and $(1.5, 0.5)$ is shown in Fig.~\ref{fg:bi2201}(b) for $\hbar\omega=4$ meV.  The smaller intensity for larger {\bf Q} is consistent with the expected fall off of the magnetic form factor at larger $|{\bf Q}|$, and therefore, suggests that the observed intensity is magnetic in origin.  Thus, the existence of spin excitations in the Bi2201 system has been demonstrated for the first time. 
 
The present success in observing spin excitations provides motivation to extend the study of spin correlations in the B2201 system.  Since both LSCO and B2201 are single-layered systems, a comparative study should yield valuable information about universal features of the spin correlations.

\section{Summary and remaining issues}
\label{sc:sum}

One of the striking features of the cuprates is that, while dynamic AF correlations tend to coexist with superconductivity, AF and SC orders generally do not coexist.  In hole-doped systems, a small density of mobile charge carriers is sufficient to destroy long-range AF order.  Remnant excitations of the AF state appear to survive at higher energies, but the correlations are reorganized at energies below \ecross.  At the same time, there is a strong damping of magnetic excitations at energies greater than that of the electronic pseudogap.  Furthermore, the magnetic spectral weight decreases monotonically with doping, disappearing in the overdoped regime together with the superconductivity.

All of these effects suggest that the doped holes and the superexchange-coupled spins organize themselves in a cooperative way to enhance both carrier mobility and locally AF correlations.  Charge and spin stripe order is certainly one motif that exhibits such cooperative self-organization.  In 214 cuprates, stripe order can be induced by suitable lattice anisotropy or by local perturbations, such as magnetic vortices or impurities.  There are indications of related nematic behavior in YBCO.  An unresolved question is: are stripes a common feature of the hole-doped cuprates?  Certainly stripe order is not common among the cuprates.  Dynamical stripes might be more common, but are there any unique signatures of fluctuating stripes?  Stripe order is only observed at modest temperatures, on the scale of $T_c$, so it cannot explain the electronic pseudogap; nevertheless, could the onset of the pseudogap reflect the self-organizing process of carriers and spins?  Neutron scattering studies of the thermal evolution of spin correlations through $T^\ast$ might help to resolve this issue, although such measurements will be challenging.

There has been a long-term debate over the nature of the magnetic excitations, and the relative importance of particle-hole excitations vs.\ the flipping of local moments.  The trend of magnetic spectral weight vs.\ doping suggests that superexchange-coupled moments likely play a dominant role over much of the phase diagram.  Particle-hole excitations must contribute at some level, but what level is that?  Is there some feature that changes with doping in a fashion that would provide circumstantial support for the role of particle-hole excitations?  The energy \ecross\ grows with doping, at least in the underdoped regime, which certainly demonstrates that the carriers and the magnetic properties are interacting, but is there a unique signature for particle-hole excitations?

Our conclusion that AF and SC orders cannot coexist is challenged by NMR experiments on cuprate families with three, four, or more CuO$_2$ layers stacked together.  The NMR measurements have been interpreted as providing evidence for the coexistence of SC and AF orders.\cite{muku08}  Now, these are complicated systems with inequivalent layers, and it can be challenging to determine with a local whether antiferromagnetic correlations are long-range commensurate or spatially modulated.  Future neutron scattering experiments could resolve this issues, if suitable crystals can be grown.

In the electron-doped cuprates, mobile charge carriers seem to be compatible with commensurate antiferromagnetic order, although superconductivity is not.  Overall, the magnetic spectral weight appears to weaken more rapidly than in the hole-doped cuprates.  Recently, Weber {\it et al.}\cite{webe10} claimed that parents of the $n$-type cuprates are not Mott insulators but Slater insulators, in which the insulating character is a consequence of the AF order. Such a system should become a metal the magnetic order is lost; however, since the cuprates are quasi-two-dimensional magnets with a large $J$, one needs to probe the system at high temperatures, $T\sim J$, to test the theory.  One would like to compare magnetic neutron scattering with transport and optical conductivity measurements.  Such a neutron scattering experiment is currently difficult 
to perform, but it is one of the challenging experiments to try in combination with new techniques such as high-energy polarized neutron spectroscopy.  Testing the nature of the parent material is also relevant to understanding the unusual ``non-doped'' superconductivity reported in thin films of  compounds such as Nd$_2$CuO$_4$ and Pr$_2$CuO$_4$.\cite{mats09}

\section{Acknowledgments}

We would like to thank K.~Hirota, H.~Kimura, M.~Kofu, S.~Iikubo, 
M.~Enoki, C.~Frost, S.-H.~Lee, Y.~Endoh, and R.~J.~Birgeneau  
for the fruitful discussions.
The work at JRR-3 and SPring-8 was partially performed 
under the Common-Use Facility Program of JAEA
and joint-research program of ISSP, the University of Tokyo.
MF is supported by Grant-in-Aid for Encouragement
of Scientific Research B (23340093).  MM and HH was supported by Grant-in-Aid for 
Encouragement of Young Scientists B (22740230)
and Scientific Research B (2234089), respectively.
JMT and GYX are supported at Brookhaven by the Office of Basic Energy Sciences, Division of Materials Science and Engineering, U.S. Department of Energy (DOE), under Contract No. DE-AC02-98CH10886.


\begin{thebibliography}{100}

\bibitem{bedn86}
J.~G. Bednorz and K.~A. M\"uller: Z. Phys. B {\bfseries 64} (1986) 189.

\bibitem{bedn88}
J.~G. Bednorz and K.~A. M\"uller: Rev. Mod. Phys. {\bfseries 60} (1988) 585.

\bibitem{ande87}
P.~W. Anderson: Science {\bfseries 235} (1987) 1169.

\bibitem{vakn87}
D.~Vaknin, S.~K. Sinha, D.~E. Moncton, D.~C. Johnston, J.~M. Newsam, C.~R.
  Safinya, and J.~H.~E.~King: Phys. Rev. Lett. {\bfseries 58} (1987) 2802.

\bibitem{scal86}
D.~J. Scalapino, E.~Loh, and J.~E. Hirsch: Phys. Rev. B {\bfseries 34} (1986)
  8190.

\bibitem{miya86}
K.~Miyake, S.~Schmitt-Rink, and C.~M. Varma: Phys. Rev. B {\bfseries 34} (1986)
  6554.

\bibitem{tran10}
J.~Tranquada and S.~M. Hayden: Neutron News {\bfseries 21} (2010) 30.

\bibitem{kast98}
M.~A. Kastner, R.~J. Birgeneau, G.~Shirane, and Y.~Endoh: Rev. Mod. Phys.
  {\bfseries 70} (1998) 897.

\bibitem{bour98}
P.~Bourges: In J.~Bok, G.~Deutscher, D.~Pavuna, and S.~A. Wolf (eds), {\em The
  Gap Symmetry and Fluctuations in High Temperature Superconductors}, 1998, p.
  349.

\bibitem{maso01}
T.~E. Mason: In J.~K.~A.~Gschneidner, L.~Eyring, and M.~B. Maple (eds), {\em
  Handbook on the Physics and Chemistry of Rare Earths, Vol.\ 31:
  High-Temperature Superconductors -- II}, 2001, pp. 281--314.

\bibitem{lynn01}
J.~W. Lynn and S.~Skanthakumar: In J.~K.~A.~Gschneidner, L.~Eyring, and M.~B.
  Maple (eds), {\em Handbood on the Physics and Chemistry of Rare Earths, Vol.
  31}, 2001, pp. 315--350.

\bibitem{tran07}
J.~M. Tranquada.
 {Neutron Scattering Studies of Antiferromagnetic Correlations in
  Cuprates}.
 In J.~R. Schrieffer and J.~S. Brooks (eds), {\em Handbook of
  High-Temperature Superconductivity}, pp. 257--298. Springer, New York, 2007.

\bibitem{kive03}
S.~A. Kivelson, I.~P. Bindloss, E.~Fradkin, V.~Oganesyan, J.~M. Tranquada,
  A.~Kapitulnik, and C.~Howald: Rev. Mod. Phys. {\bfseries 75} (2003) 1201.

\bibitem{deml04}
E.~Demler, W.~Hanke, and S.-C. Zhang: Rev. Mod. Phys. {\bfseries 76} (2004)
  909.

\bibitem{lee06}
P.~A. Lee, N.~Nagaosa, and X.-G. Wen: Rev. Mod. Phys. {\bfseries 78} (2006) 17.

\bibitem{esch06}
M.~Eschrig: Adv. Phys. {\bfseries 55} (2006) 47.

\bibitem{ogat08}
M.~Ogata and H.~Fukuyama: Rep. Prog. Phys. {\bfseries 71} (2008) 036501.

\bibitem{birg06}
R.~J. Birgeneau, C.~Stock, J.~M. Tranquada, and K.~Yamada: J. Phys. Soc. Jpn.
  {\bfseries 75} (2006) 111003.

\bibitem{hayd91}
S.~M. Hayden, G.~Aeppli, R.~Osborn, A.~D. Taylor, T.~G. Perring, S.-W. Cheong,
  and Z.~Fisk: Phys. Rev. Lett. {\bfseries 67} (1991) 3622.

\bibitem{hayd96a}
S.~M. Hayden, G.~Aeppli, H.~A. Mook, T.~G. Perring, T.~E. Mason, S.-W. Cheong,
  and Z.~Fisk: Phys. Rev. Lett. {\bfseries 76} (1996) 1344.

\bibitem{cold01}
R.~Coldea, S.~M. Hayden, G.~Aeppli, T.~G. Perring, C.~D. Frost, T.~E. Mason,
  S.-W. Cheong, and Z.~Fisk: Phys. Rev. Lett. {\bfseries 86} (2001) 5377.

\bibitem{head10}
N.~S. Headings, S.~M. Hayden, R.~Coldea, and T.~G. Perring: Phys. Rev. Lett.
  {\bfseries 105} (2010) 247001.

\bibitem{dai01}
P.~Dai, H.~A. Mook, R.~D. Hunt, and F.~Do\u{g}an: Phys. Rev. B {\bfseries 63}
  (2001) 054525.

\bibitem{chri04}
N.~B. Christensen, D.~F. McMorrow, H.~M. R{\o}nnow, B.~Lake, S.~M. Hayden,
  G.~Aeppli, T.~G. Perring, M.~Mangkorntong, M.~Nohara, and H.~Tagaki: Phys.
  Rev. Lett. {\bfseries 93} (2004) 147002.

\bibitem{tran04b}
J.~M. Tranquada, C.~H. Lee, K.~Yamada, Y.~S. Lee, L.~P. Regnault, and H.~M.
  R{\o}nnow: Phys. Rev. B {\bfseries 69} (2004) 174507.

\bibitem{pail06}
S.~Pailh\`es, C.~Ulrich, B.~Fauqu\'e, V.~Hinkov, Y.~Sidis, A.~Ivanov, C.~T.
  Lin, B.~Keimer, and P.~Bourges: Phys. Rev. Lett. {\bfseries 96} (2006)
  257001.

\bibitem{yu09}
G.~Yu, Y.~Li, E.~M. Motoyama, and M.~Greven: Nat. Phys. {\bfseries 5} (2009)
  873.

\bibitem{hayd04}
S.~M. Hayden, H.~A. Mook, P.~Dai, T.~G. Perring, and F.~Do\u{g}an: Nature
  {\bfseries 429} (2004) 531.

\bibitem{tran04}
J.~M. Tranquada, H.~Woo, T.~G. Perring, H.~Goka, G.~D. Gu, G.~Xu, M.~Fujita,
  and K.~Yamada: Nature {\bfseries 429} (2004) 534.

\bibitem{stoc05}
C.~Stock, W.~J.~L. Buyers, R.~A. Cowley, P.~S. Clegg, R.~Coldea, C.~D. Frost,
  R.~Liang, D.~Peets, D.~Bonn, W.~N. Hardy, and R.~J. Birgeneau: Phys. Rev. B
  {\bfseries 71} (2005) 024522.

\bibitem{stoc10}
C.~Stock, R.~A. Cowley, W.~J.~L. Buyers, C.~D. Frost, J.~W. Taylor, D.~Peets,
  R.~Liang, D.~Bonn, and W.~N. Hardy: Phys. Rev. B {\bfseries 82} (2010)
  174505.

\bibitem{vign07}
B.~Vignolle, S.~M. Hayden, D.~F. McMorrow, H.~M. {R{\o}nnow}, B.~Lake, C.~D.
  Frost, and T.~G. Perring: Nat. Phys. {\bfseries 3} (2007) 163.

\bibitem{kofu07}
M.~Kofu, T.~Yokoo, F.~Trouw, and K.~Yamada: http://arXiv:0710.5766 (2007).

\bibitem{hink07}
V.~Hinkov, P.~Bourges, S.~Pailhes, Y.~Sidis, A.~Ivanov, C.~D. Frost, T.~G.
  Perring, C.~T. Lin, D.~P. Chen, and B.~Keimer: Nat. Phys. {\bfseries 3}
  (2007) 780.

\bibitem{lips09}
O.~J. Lipscombe, B.~Vignolle, T.~G. Perring, C.~D. Frost, and S.~M. Hayden:
  Phys. Rev. Lett. {\bfseries 102} (2009) 167002.

\bibitem{hink04}
V.~Hinkov, S.~Pailh\`{e}s, P.~Bourges, Y.~Sidis, A.~Ivanov, A.~Kulakov, C.~T.
  Lin, D.~P. Chen, C.~Bernhard, and B.~Keimer: Nature {\bfseries 430} (2004)
  650.

\bibitem{hink08}
V.~Hinkov, D.~Haug, B.~Fauqu\'e, P.~Bourges, Y.~Sidis, A.~Ivanov, C.~Bernhard,
  C.~T. Lin, and B.~Keimer: Science {\bfseries 319} (2008) 597.

\bibitem{haug10}
D.~Haug, V.~Hinkov, Y.~Sidis, P.~Bourges, N.~B. Christensen, A.~Ivanov,
  T.~Keller, C.~T. Lin, and B.~Keimer: New J. Phys. {\bfseries 12} (2010)
  105006.

\bibitem{hink10}
V.~Hinkov, C.~Lin, M.~Raichle, B.~Keimer, Y.~Sidis, P.~Bourges, S.~Pailh{\`e}s,
  and A.~Ivanov: Eur. Phys. J. Special Topics {\bfseries 188} (2010) 113.

\bibitem{xu09}
G.~Xu, G.~D. Gu, M.~Hucker, B.~Fauque, T.~G. Perring, L.~P. Regnault, and J.~M.
  Tranquada: Nat. Phys. {\bfseries 5} (2009) 642.

\bibitem{suga03}
S.~Sugai, H.~Suzuki, Y.~Takayanagi, T.~Hosokawa, and N.~Hayamizu: Phys. Rev. B
  {\bfseries 68} (2003) 184504.

\bibitem{fong99}
H.~F. Fong, P.~Bourges, Y.~Sidis, L.~P. Regnault, A.~Ivanov, G.~D. Gu,
  N.~Koshizuka, and B.~Keimer: Nature {\bfseries 398} (1999) 588.

\bibitem{he01}
H.~He, Y.~Sidis, P.~Bourges, G.~D. Gu, A.~Ivanov, N.~Koshizuka, B.~Liang, C.~T.
  Lin, L.~P. Regnault, E.~Schoenherr, and B.~Keimer: Phys. Rev. Lett.
  {\bfseries 86} (2001) 1610.

\bibitem{capo07}
L.~Capogna, B.~Fauque, Y.~Sidis, C.~Ulrich, P.~Bourges, S.~Pailhes, A.~Ivanov,
  J.~L. Tallon, B.~Liang, C.~T. Lin, A.~I. Rykov, and B.~Keimer: Phys. Rev. B
  {\bfseries 75} (2007) 060502.

\bibitem{mats09c}
M.~Matsuura, Y.~Yoshida, H.~Eisaki, N.~Kaneko, C.-H. Lee, and K.~Hirota: J.
  Phys. Soc. Jpn. {\bfseries 78} (2009) 074703.

\bibitem{fauq07}
B.~{Fauqu\'e}, Y.~Sidis, L.~Capogna, A.~Ivanov, K.~Hradil, C.~Ulrich, A.~I.
  Rykov, B.~Keimer, and P.~Bourges: Phys. Rev. B {\bfseries 76} (2007) 214512.

\bibitem{wen08}
J.~S. Wen, Z.~J. Xu, G.~Y. Xu, M.~{H\"ucker}, J.~M. Tranquada, and G.~Gu: J.
  Cryst. Growth {\bfseries 310} (2008) 1401.

\bibitem{lu92}
J.~P. Lu: Phys. Rev. Lett. {\bfseries 68} (1992) 125.

\bibitem{bulu93}
N.~Bulut and D.~J. Scalapino: Phys. Rev. B {\bfseries 47} (1993) 3419.

\bibitem{zha93}
Y.~Zha, K.~Levin, and Q.~Si: Phys. Rev. B {\bfseries 47} (1993) 9124.

\bibitem{norm07}
M.~R. Norman: Phys. Rev. B {\bfseries 75} (2007) 184514.

\bibitem{breh10}
{Brehm, S.}, {Arrigoni, E.}, {Aichhorn, M.}, and {Hanke, W.}: Europhys. Lett.
  {\bfseries 89} (2010) 27005.

\bibitem{lake99}
B.~Lake, G.~Aeppli, T.~E. Mason, A.~Schr\"oder, D.~F. McMorrow, K.~Lefmann,
  M.~Isshiki, M.~Nohara, H.~Takagi, and S.~M. Hayden: Nature {\bfseries 400}
  (1999) 43.

\bibitem{mats08}
M.~Matsuda, M.~Fujita, S.~Wakimoto, J.~A. Fernandez-Baca, J.~M. Tranquada, and
  K.~Yamada: Phys. Rev. Lett. {\bfseries 101} (2008) 197001.

\bibitem{sun10}
K.~Sun, M.~J. Lawler, and E.-A. Kim: Phys. Rev. Lett. {\bfseries 104} (2010)
  106405.

\bibitem{goka03}
H.~Goka, S.~Kuroshima, M.~Fujita, K.~Yamada, H.~Hiraka, Y.~Endoh, and C.~D.
  Frost: Physica C {\bfseries 388--389} (2003) 239.

\bibitem{hira11}
H.~Hiraka, H.~Goka, M.~Fujita, C.~D. Frost, M.~Matsuda, K.~Ikeuchi, A.~Komarek,
  M.~Braden, J.~M. Tranquada, and K.~Yamada: (unpublished) .

\bibitem{bati01}
C.~D. Batista, G.~Ortiz, and A.~V. Balatsky: Phys. Rev. B {\bfseries 64} (2001)
  172508.

\bibitem{krug03}
F.~Kr\"uger and S.~Scheidl: Phys. Rev. B {\bfseries 67} (2003) 134512.

\bibitem{mats00}
M.~Matsuda, Y.~S. Lee, M.~Greven, M.~A. Kastner, R.~J. Birgeneau, K.~Yamada,
  Y.~Endoh, P.~B\"oni, S.-H. Lee, S.~Wakimoto, and G.~Shirane: Phys. Rev. B
  {\bfseries 61} (2000) 4326.

\bibitem{hira01}
H.~Hiraka, Y.~Endoh, M.~Fujita, Y.~S. Lee, J.~Kulda, A.~Ivanov, and R.~J.
  Birgeneau: J. Phys. Soc. Jpn {\bfseries 70} (2001) 853.

\bibitem{gran04}
M.~Granath: Phys. Rev. B {\bfseries 69} (2004) 214433.

\bibitem{yosh06}
T.~{Yoshida {\it et al.}}: Phys. Rev. B {\bfseries 74} (2006) 224510.

\bibitem{seib06}
G.~Seibold and J.~Lorenzana: Phys. Rev. B {\bfseries 73} (2006) 144515.

\bibitem{seib09}
G.~Seibold and J.~Lorenzana: Phys. Rev. B {\bfseries 80} (2009) 012509.

\bibitem{carl04}
E.~W. Carlson, D.~X. Yao, and D.~K. Campbell: Phys. Rev. B {\bfseries 70}
  (2004) 064505.

\bibitem{bour03}
P.~Bourges, Y.~Sidis, M.~Braden, K.~Nakajima, and J.~M. Tranquada: Phys. Rev.
  Lett. {\bfseries 90} (2003) 147202.

\bibitem{woo05}
H.~Woo, A.~T. Boothroyd, K.~Nakajima, T.~G. Perring, C.~D. Frost, P.~G.
  Freeman, D.~Prabhakaran, K.~Yamada, and J.~M. Tranquada: Phys. Rev. B
  {\bfseries 72} (2005) 064437.

\bibitem{hufn08}
S.~{H\"{u}fner}, M.~A. Hossain, A.~Damascelli, and G.~A. Sawatzky: Rep. Prog.
  Phys. {\bfseries 71} (2008) 062501.

\bibitem{gork06}
L.~P. {Gor'kov} and G.~B. {Teitel'baum}: Phys. Rev. Lett. {\bfseries 97} (2006)
  247003.

\bibitem{ando04}
Y.~Ando, Y.~Kurita, S.~Komiya, S.~Ono, and K.~Segawa: Phys. Rev. Lett.
  {\bfseries 92} (2004) 197001.

\bibitem{lee05}
Y.~S. Lee, K.~Segawa, Z.~Q. Li, W.~J. Padilla, M.~Dumm, S.~V. Dordevic, C.~C.
  Homes, Y.~Ando, and D.~N. Basov: Phys. Rev. B {\bfseries 72} (2005) 054529.

\bibitem{hayd96b}
S.~M. Hayden, G.~Aeppli, T.~G. Perring, H.~A. Mook, and F.~{Do\u gan}: Phys.
  Rev. B {\bfseries 54} (1996) R6905.

\bibitem{ande97}
P.~W. Anderson: Adv. Phys. {\bfseries 46} (1997) 3.

\bibitem{dama03}
A.~Damascelli, Z.-X. Shen, and Z.~Hussain: Rev. Mod. Phys. {\bfseries 75}
  (2003) 473.

\bibitem{trug88}
S.~A. Trugman: Phys. Rev. B {\bfseries 37} (1988) 1597.

\bibitem{lau11}
B.~Lau, M.~Berciu, and G.~A. Sawatzky: Phys. Rev. Lett. {\bfseries 106} (2011)
  036401.

\bibitem{idet11}
S.~Ideta, T.~Yoshida, A.~Fujimori, H.~Anzai, T.~Fujita, A.~Ino, M.~Arita,
  H.~Namatame, M.~Taniguchi, Z.-X. Shen, K.~Takashima, K.~Kojima, and
  S.~Uchida: arXiv:1104.0313v1 (2011).

\bibitem{waki07b}
S.~Wakimoto, K.~Yamada, J.~M. Tranquada, C.~D. Frost, R.~J. Birgeneau, and
  H.~Zhang: Phys. Rev. Lett. {\bfseries 98} (2007) 247003.

\bibitem{lips07}
O.~J. Lipscombe, S.~M. Hayden, B.~Vignolle, D.~F. McMorrow, and T.~G. Perring:
  Phys. Rev. Lett. {\bfseries 99} (2007) 067002.

\bibitem{regn95}
L.~P. Regnault, P.~Bourges, P.~Burlet, J.~Y. Henry, J.~Rossat-Mignod, Y.~Sidis,
  and C.~Vettier: Physica B {\bfseries 213\&214} (1995) 48.

\bibitem{rezn08}
D.~Reznik, J.-P. Ismer, I.~Eremin, L.~Pintschovius, T.~Wolf, M.~Arai, Y.~Endoh,
  T.~Masui, and S.~Tajima: Phys. Rev. B {\bfseries 78} (2008) 132503.

\bibitem{leta11}
M.~Le~Tacon, G.~Ghiringhelli, J.~Chaloupka, M.~M. Sala, V.~Hinkov, M.~W.
  Haverkort, M.~Minola, M.~Bakr, K.~J. Zhou, S.~Blanco-Canosa, C.~Monney, Y.~T.
  Song, G.~L. Sun, C.~T. Lin, G.~M. De~Luca, M.~Salluzzo, G.~Khaliullin,
  T.~Schmitt, L.~Braicovich, and B.~Keimer: Nat. Phys. {\bfseries 7} (2011)
  725.

\bibitem{uemu03}
Y.~J. Uemura: Solid State Communications {\bfseries 126} (2003) 23.

\bibitem{tran95a}
J.~M. Tranquada, B.~J. Sternlieb, J.~D. Axe, Y.~Nakamura, and S.~Uchida: Nature
  {\bfseries 375} (1995) 561.

\bibitem{fuji04}
M.~Fujita, H.~Goka, K.~Yamada, J.~M. Tranquada, and L.~P. Regnault: Phys. Rev.
  B {\bfseries 70} (2004) 104517.

\bibitem{vojt09}
M.~Vojta: Adv. Phys. {\bfseries 58} (2009) 699.

\bibitem{li07}
Q.~Li, M.~{H\"ucker}, G.~D. Gu, A.~M. Tsvelik, and J.~M. Tranquada: Phys. Rev.
  Lett. {\bfseries 99} (2007) 067001.

\bibitem{tran08}
J.~M. Tranquada, G.~D. Gu, M.~H{\"u}cker, Q.~Jie, H.-J. Kang, R.~Klingeler,
  Q.~Li, N.~Tristan, J.~S. Wen, G.~Y. Xu, Z.~J. Xu, J.~Zhou, and
  M.~v.~Zimmermann: Phys. Rev. B {\bfseries 78} (2008) 174529.

\bibitem{hime02}
A.~Himeda, T.~Kato, and M.~Ogata: Phys. Rev. Lett. {\bfseries 88} (2002)
  117001.

\bibitem{berg07}
E.~Berg, E.~Fradkin, E.-A. Kim, S.~A. Kivelson, V.~Oganesyan, J.~M. Tranquada,
  and S.~C. Zhang: Phys. Rev. Lett. {\bfseries 99} (2007) 127003.

\bibitem{berg09b}
E.~Berg, E.~Fradkin, S.~A. Kivelson, and J.~M. Tranquada: New J. Phys.
  {\bfseries 11} (2009) 115004.

\bibitem{huck11}
M.~H\"ucker, M.~v.~Zimmermann, G.~D. Gu, Z.~J. Xu, J.~S. Wen, G.~Xu, H.~J.
  Kang, A.~Zheludev, and J.~M. Tranquada: Phys. Rev. B {\bfseries 83} (2011)
  104506.

\bibitem{fuji06}
M.~Fujita, M.~Matsuda, H.~Goka, T.~Adachi, Y.~Koike, and K.~Yamada: J. Phys.
  Conf. Ser. {\bfseries 51} (2006) 510.

\bibitem{kim08}
Y.-J. Kim, G.~D. Gu, T.~Gog, and D.~Casa: Phys. Rev. B {\bfseries 77} (2008)
  064520.

\bibitem{duns08}
S.~R. Dunsiger, Y.~Zhao, Z.~Yamani, W.~J.~L. Buyers, H.~Dabkowska, and B.~D.
  Gaulin: Phys. Rev. B {\bfseries 77} (2008) 224410.

\bibitem{duns08b}
S.~R. Dunsiger, Y.~Zhao, B.~D. Gaulin, Y.~Qiu, P.~Bourges, Y.~Sidis, J.~R.~D.
  Copley, A.~Kallin, E.~M. Mazurek, and H.~A. Dabkowska: Phys. Rev. B
  {\bfseries 78} (2008) 092507.

\bibitem{zhao07}
Y.~Zhao, B.~D. Gaulin, J.~P. Castellan, J.~P.~C. Ruff, S.~R. Dunsiger, G.~D.
  Gu, and H.~A. Dabkowska: Phys. Rev. B {\bfseries 76} (2007) 184121.

\bibitem{abba05}
P.~Abbamonte, A.~Rusydi, S.~Smadici, G.~D. Gu, G.~A. Sawatzky, and D.~L. Feng:
  Nat. Phys. {\bfseries 1} (2005) 155.

\bibitem{fink11}
J.~Fink, V.~Soltwisch, J.~Geck, E.~Schierle, E.~Weschke, and B.~B\"uchner:
  Phys. Rev. B {\bfseries 83} (2011) 092503.

\bibitem{klau00}
H.-H. Klauss, W.~Wagener, M.~Hillberg, W.~Kopmann, H.~Walf, F.~J. Litterst,
  M.~H\"ucker, and B.~B\"uchner: Phys. Rev. Lett. {\bfseries 85} (2000) 4590.

\bibitem{huck07}
M.~{H\"ucker}, G.~D. Gu, J.~M. Tranquada, M.~v.~Zimmermann, H.-H. Klauss, N.~J.
  Curro, M.~Braden, and B.~{B\"uchner}: Physica C {\bfseries 460--462} (2007)
  170.

\bibitem{hoff02}
J.~E. Hoffman, E.~W. Hudson, K.~M. Lang, V.~Madhavan, H.~Eisaki, S.~Uchida, and
  J.~C. Davis: Science {\bfseries 295} (2002) 466.

\bibitem{hana04}
T.~Hanaguri, C.~Lupien, Y.~Kohsaka, D.-H. Lee, M.~Azuma, M.~Takano, H.~Takagi,
  and J.~C. Davis: Nature {\bfseries 430} (2004) 1001.

\bibitem{vojt02}
M.~Vojta: Phys. Rev. B {\bfseries 66} (2002) 104505.

\bibitem{fine04}
B.~V. Fine: Phys. Rev. B {\bfseries 70} (2004) 224508.

\bibitem{sush04}
O.~P. Sushkov and V.~N. Kotov: Phys. Rev. B {\bfseries 70} (2004) 024503.

\bibitem{chri07}
N.~B. Christensen, H.~M. {R{\o}nnow}, J.~Mesot, R.~A. Ewings, N.~Momono,
  M.~Oda, M.~Ido, M.~Enderle, D.~F. McMorrow, and A.~T. Boothroyd: Phys. Rev.
  Lett. {\bfseries 98} (2007) 197003.

\bibitem{kohs07}
Y.~Kohsaka, C.~Taylor, K.~Fujita, A.~Schmidt, C.~Lupien, T.~Hanaguri, M.~Azuma,
  M.~Takano, H.~Eisaki, H.~Takagi, S.~Uchida, and J.~C. Davis: Science
  {\bfseries 315} (2007) 1380.

\bibitem{lawl10}
M.~J. Lawler, K.~Fujita, J.~Lee, A.~R. Schmidt, Y.~Kohsaka, C.~K. Kim,
  H.~Eisaki, S.~Uchida, J.~C. Davis, J.~P. Sethna, and E.-A. Kim: Nature
  {\bfseries 466} (2010) 347.

\bibitem{park10}
C.~V. Parker, P.~Aynajian, E.~H. da~Silva~Neto, A.~Pushp, S.~Ono, J.~Wen,
  Z.~Xu, G.~Gu, and A.~Yazdani: Nature {\bfseries 468} (2010) 677.

\bibitem{taji01}
S.~Tajima, T.~Noda, H.~Eisaki, and S.~Uchida: Phys. Rev. Lett. {\bfseries 86}
  (2001) 500.

\bibitem{baru08}
S.~Baruch and D.~Orgad: Phys. Rev. B {\bfseries 77} (2008) 174502.

\bibitem{vall06}
T.~Valla, A.~V. Federov, J.~Lee, J.~C. Davis, and G.~D. Gu: Science {\bfseries
  314} (2006) 1914.

\bibitem{he09}
R.-H. He, K.~Tanaka, S.-K. Mo, T.~Sasagawa, M.~Fujita, T.~Adachi, N.~Mannella,
  K.~Yamada, Y.~Koike, Z.~Hussain, and Z.-X. Shen: Nat. Phys. {\bfseries 5}
  (2009) 119.

\bibitem{stoc08}
C.~Stock, W.~J.~L. Buyers, Z.~Yamani, Z.~Tun, R.~J. Birgeneau, R.~Liang,
  D.~Bonn, and W.~N. Hardy: Phys. Rev. B {\bfseries 77} (2008) 104513.

\bibitem{such10}
A.~Suchaneck, V.~Hinkov, D.~Haug, L.~Schulz, C.~Bernhard, A.~Ivanov, K.~Hradil,
  C.~T. Lin, P.~Bourges, B.~Keimer, and Y.~Sidis: Phys. Rev. Lett. {\bfseries
  105} (2010) 037207.

\bibitem{kive98}
S.~A. Kivelson, E.~Fradkin, and V.~J. Emery: Nature {\bfseries 393} (1998) 550.

\bibitem{daou10}
R.~Daou, J.~Chang, D.~LeBoeuf, O.~Cyr-Choiniere, F.~Laliberte,
  N.~Doiron-Leyraud, B.~J. Ramshaw, R.~Liang, D.~A. Bonn, W.~N. Hardy, and
  L.~Taillefer: Nature {\bfseries 463} (2010) 519.

\bibitem{dubr11}
A.~Dubroka, M.~R\"ossle, K.~W. Kim, V.~K. Malik, D.~Munzar, D.~N. Basov, A.~A.
  Schafgans, S.~J. Moon, C.~T. Lin, D.~Haug, V.~Hinkov, B.~Keimer, T.~Wolf,
  J.~G. Storey, J.~L. Tallon, and C.~Bernhard: Phys. Rev. Lett. {\bfseries 106}
  (2011) 047006.

\bibitem{lali11}
F.~Lalibert\'e, J.~Chang, N.~Doiron-Leyraud, E.~Hassinger, R.~Daou, M.~Rondeau,
  B.~J. Ramshaw, R.~Liang, D.~A. Bonn, W.~N. Hardy, S.~Pyon, T.~Takayama,
  H.~Takagi, I.~Sheikin, L.~Malone, C.~Proust, K.~Behnia, and L.~Taillefer:
  Nat. Commun. {\bfseries 2} (2011) 432.

\bibitem{lebo11}
D.~LeBoeuf, N.~Doiron-Leyraud, B.~Vignolle, M.~Sutherland, B.~J. Ramshaw,
  J.~Levallois, R.~Daou, F.~Lalibert\'e, O.~Cyr-Choini\`ere, J.~Chang, Y.~J.
  Jo, L.~Balicas, R.~Liang, D.~A. Bonn, W.~N. Hardy, C.~Proust, and
  L.~Taillefer: Phys. Rev. B {\bfseries 83} (2011) 054506.

\bibitem{boot11}
A.~T. Boothroyd, P.~Babkevich, D.~Prabhakaran, and P.~G. Freeman: Nature
  {\bfseries 471} (2011) 341.

\bibitem{kata00}
S.~Katano, M.~Sato, K.~Yamada, T.~Suzuki, and T.~Fukase: Phys. Rev. B
  {\bfseries 62} (2000) R14677.

\bibitem{lake02}
B.~Lake, H.~M. {R\o nnow}, N.~B. Christensen, G.~Aeppli, K.~Lefmann, D.~F.
  McMorrow, P.~Vorderwisch, P.~Smeibidl, N.~Mangkorntong, T.~Sasagawa,
  M.~Nohara, H.~Takagi, and T.~E. Mason: Nature {\bfseries 415} (2002) 299.

\bibitem{chan08}
J.~Chang, C.~Niedermayer, R.~Gilardi, N.~Christensen, H.~Ronnow, D.~McMorrow,
  M.~Ay, J.~Stahn, O.~Sobolev, A.~Hiess, S.~Pailhes, C.~Baines, N.~Momono,
  M.~Oda, M.~Ido, and J.~Mesot: Phys. Rev. B {\bfseries 78} (2008) 104525.

\bibitem{khay02}
B.~Khaykovich, Y.~S. Lee, R.~W. Erwin, S.-H. Lee, S.~Wakimoto, K.~J. Thomas,
  M.~A. Kastner, and R.~J. Birgeneau: Phys. Rev. B {\bfseries 66} (2002)
  014528.

\bibitem{khay03}
B.~Khaykovich, R.~J. Birgeneau, F.~C. Chou, R.~W. Erwin, M.~A. Kastner, S.-H.
  Lee, Y.~S. Lee, P.~Smeibidl, P.~Vorderwisch, and S.~Wakimoto: Phys. Rev. B
  {\bfseries 67} (2003) 054501.

\bibitem{wen08b}
J.~S. Wen, Z.~X. Xu, G.~Y. Xu, J.~M. Tranquada, G.~D. Gu, S.~Chang, and H.~J.
  Kang: Phys. Rev. B {\bfseries 78} (2008) 212506.

\bibitem{waki03}
S.~Wakimoto, R.~J. Birgeneau, Y.~Fujimaki, N.~Ichikawa, T.~Kasuga, Y.~J. Kim,
  K.~M. Kojima, S.-H. Lee, H.~Niko, J.~M. Tranquada, S.~Uchida, and
  M.~v.~Zimmermann: Phys. Rev. B {\bfseries 67} (2003) 184419.

\bibitem{deml01}
E.~Demler, S.~Sachdev, and Y.~Zhang: Phys. Rev. Lett. {\bfseries 87} (2001)
  067202.

\bibitem{kive02}
S.~A. Kivelson, D.-H. Lee, E.~Fradkin, and V.~Oganesyan: Phys. Rev. B
  {\bfseries 66} (2002) 144516.

\bibitem{lake01}
B.~Lake, G.~Aeppli, K.~N. Clausen, D.~F. McMorrow, K.~Lefmann, N.~E. Hussey,
  N.~Mangkorntong, M.~Nohara, H.~Takagi, T.~E. Mason, and A.~Schroder: Science
  {\bfseries 291} (2001) 1759.

\bibitem{gila04}
R.~Gilardi, A.~Hiess, N.~Momono, M.~Oda, M.~Ido, and J.~Mesot: Europhys. Lett.
  {\bfseries 66} (2004) 840.

\bibitem{chan09}
J.~Chang, N.~B. Christensen, C.~Niedermayer, K.~Lefmann, H.~M. R\o{}nnow, D.~F.
  McMorrow, A.~Schneidewind, P.~Link, A.~Hiess, M.~Boehm, R.~Mottl,
  S.~Pailh\'es, N.~Momono, M.~Oda, M.~Ido, and J.~Mesot: Phys. Rev. Lett.
  {\bfseries 102} (2009) 177006.

\bibitem{khay05}
B.~Khaykovich, S.~Wakimoto, R.~J. Birgeneau, M.~A. Kastner, Y.~S. Lee,
  P.~Smeibidl, P.~Vorderwisch, and K.~Yamada: Phys. Rev. B {\bfseries 71}
  (2005) 220508(R).

\bibitem{wen11}
J.~Wen, Q.~Jie, Q.~Li, M.~H{\"u}cker, M.~v.~Zimmermann, S.~J. Han, Z.~Xu, D.~K.
  Singh, L.~Zhang, G.~Gu, and J.~M. Tranquada: arXiv:1009.0031v3 (2010).

\bibitem{haug09}
D.~Haug, V.~Hinkov, A.~Suchaneck, D.~S. Inosov, N.~B. Christensen,
  C.~Niedermayer, P.~Bourges, Y.~Sidis, J.~T. Park, A.~Ivanov, C.~T. Lin,
  J.~Mesot, and B.~Keimer: Phys. Rev. Lett. {\bfseries 103} (2009) 017001.

\bibitem{stoc09}
C.~Stock, W.~J.~L. Buyers, K.~C. Rule, J.-H. Chung, R.~Liang, D.~Bonn, and
  W.~N. Hardy: Phys. Rev. B {\bfseries 79} (2009) 184514.

\bibitem{nach96}
B.~Nachumi, A.~Keren, K.~Kojima, M.~Larkin, G.~M. Luke, J.~Merrin,
  O.~Tchernysh\"ov, Y.~J. Uemura, N.~Ichikawa, M.~Goto, and S.~Uchida: Phys.
  Rev. Lett. {\bfseries 77} (1996) 5421.

\bibitem{pan00}
S.~H. Pan, E.~W. Hudson, K.~M. Lang, H.~Eisaki, S.~Uchida, and J.~C. Davis:
  Nature {\bfseries 403} (2000) 746.

\bibitem{huds01}
E.~W. Hudson, K.~M. Lang, V.~Madhavan, S.~H. Pan, H.~Eisaki, S.~Uchida, and
  J.~C. Davis: Nature {\bfseries 411} (2001) 920.

\bibitem{hiro98}
K.~Hirota, K.~Yamada, I.~Tanaka, and H.~Kojima: Physica B {\bfseries 241--243}
  (1998) 817.

\bibitem{kimu03b}
H.~Kimura, M.~Kofu, Y.~Matsumoto, and K.~Hirota: Phys. Rev. Lett. {\bfseries
  91} (2003) 067002.

\bibitem{kofu05}
M.~Kofu, H.~Kimura, and K.~Hirota: Phys. Rev. B {\bfseries 72} (2005) 064502.

\bibitem{toku97}
Y.~Tokunaga, K.~Ishida, Y.~Kitaoka, and K.~Asayama: Solid State Commun.
  {\bfseries 103} (1997) 43.

\bibitem{mats09}
M.~Matsuura, M.~Kofu, H.~Kimura, and K.~Hirota: J. Phys. Soc. Jpn. {\bfseries
  78} (2009) 114703.

\bibitem{bhat92}
V.~Bhat, C.~N.~R. Rao, and J.~M. Honig: Physica C {\bfseries 191} (1992) 271.

\bibitem{mend94}
P.~Mendels, H.~Alloul, G.~Collin, N.~Blanchard, J.~F. Marucco, and J.~Bobroff:
  Physica C {\bfseries 235--240} (1994) 1595.

\bibitem{naka98c}
T.~Nakano, N.~Momono, T.~Nagata, M.~Oda, and M.~Ido: Phys. Rev. B {\bfseries
  58} (1998) 5831.

\bibitem{wata03}
N.~Watanabe, T.~Masui, Y.~Itoh, T.~Machi, I.~Kato, N.~Koshizuka, and
  M.~Murakami: Physica C {\bfseries 388--389} (2003) 241.

\bibitem{hira05}
H.~Hiraka, T.~Machi, N.~Watanabe, Y.~Itoh, M.~Matsuda, and K.~Yamada: J. Phys.
  Soc. Jpn. {\bfseries 74} (2005) 2197.

\bibitem{mats06}
M.~Matsuda, M.~Fujita, and K.~Yamada: Phys. Rev. B {\bfseries 73} (2006)
  140503(R).

\bibitem{hira07}
H.~Hiraka, S.~Ohta, S.~Wakimoto, M.~Matsuda, and K.~Yamada: J. Phys. Soc. Jpn.
  {\bfseries 76} (2007) 074703.

\bibitem{hira08}
H.~Hiraka, S.~Wakimoto, M.~Matsuda, D.~Matsumura, Y.~Nishihata, J.~Mizuki, and
  K.~Yamada: J. Phys. Chem. Solids {\bfseries 69} (2008) 3136.

\bibitem{hira09}
H.~Hiraka, D.~Matsumura, Y.~Nishihata, J.~Mizuki, and K.~Yamada: Phys. Rev.
  Lett. {\bfseries 102} (2009) 037002.

\bibitem{tsut09}
K.~Tsutsui, A.~Toyama, T.~Tohyama, and S.~Maekawa: Phys. Rev. B {\bfseries 80}
  (2009) 224519.

\bibitem{tana10}
Y.~Tanabe, K.~Suzuki, T.~Adachi, Y.~Koike, T.~Kawamata, R.~Risdiana, T.~Suzuki,
  and I.~Watanabe: J. Phys. Soc. Jpn. {\bfseries 79} (2010) 023706.

\bibitem{huck10}
M.~H\"ucker, M.~v.~Zimmermann, M.~Debessai, J.~S. Schilling, J.~M. Tranquada,
  and G.~D. Gu: Phys. Rev. Lett. {\bfseries 104} (2010) 057004.

\bibitem{fuji02}
M.~Fujita, H.~Goka, K.~Yamada, and M.~Matsuda: Phys. Rev. Lett. {\bfseries 88}
  (2002) 167008.

\bibitem{kimu04}
H.~Kimura, Y.~Noda, H.~Goka, M.~Fujita, K.~Yamada, M.~Mizumaki, N.~Ikeda, and
  H.~Ohsumi: Phys. Rev. B {\bfseries 70} (2004) 134512.

\bibitem{fuji08b}
M.~Fujita, M.~Enoki, and K.~Yamada: J. Phys. Chem. Solids {\bfseries 69} (2008)
  3167.

\bibitem{fuji09a}
M.~Fujita, M.~Enoki, S.~Iikubo, and K.~Yamada: J. Supercond. Nov. Magn.
  {\bfseries 22} (2009) 243.

\bibitem{fuji09b}
M.~Fujita, M.~Enoki, S.~Iikubo, K.~Kudo, N.~Kobayashi, and K.~Yamada:
  arXiv:0903.5391 (2009).

\bibitem{hira10}
H.~Hiraka, Y.~Hayashi, S.~Wakimoto, M.~Takeda, K.~Kakurai, T.~Adachi, Y.~Koike,
  I.~Yamada, M.~Miyazaki, M.~Hiraishi, S.~Takeshita, A.~Kohda, R.~Kadono, J.~M.
  Tranquada, and K.~Yamada: Phys. Rev. B {\bfseries 81} (2010) 144501.

\bibitem{waki10}
S.~Wakimoto, H.~Hiraka, K.~Kudo, D.~Okamoto, T.~Nishizaki, K.~Kakurai, T.~Hong,
  A.~Zheludev, J.~M. Tranquada, N.~Kobayashi, and K.~Yamada: Phys. Rev. B
  {\bfseries 82} (2010) 064507.

\bibitem{he11}
R.-H. He, M.~Fujita, M.~Enoki, M.~Hashimoto, S.~Iikubo, S.-K. Mo, H.~Yao,
  T.~Adachi, Y.~Koike, Z.~Hussain, Z.-X. Shen, and K.~Yamada: Phys. Rev. Lett.
  {\bfseries 107} (2011) 127002.

\bibitem{nuck89}
N.~N\"ucker, H.~Romberg, X.~X. Xi, J.~Fink, B.~Gegenheimer, and Z.~X. Zhao:
  Phys. Rev. B {\bfseries 39} (1989) 6619.

\bibitem{bian87}
A.~Bianconi, A.~C. Castellano, M.~D. Santis, P.~Rudolf, P.~Lagarde, A.~M.
  Flank, and A.~Marcelli: Solid State Communications {\bfseries 63} (1987)
  1009.

\bibitem{chen91}
C.~T. Chen, F.~Sette, Y.~Ma, M.~S. Hybertsen, E.~B. Stechel, W.~M.~C. Foulkes,
  M.~Schulter, S.-W. Cheong, A.~S. Cooper, L.~W. Rupp, B.~Batlogg, Y.~L. Soo,
  Z.~H. Ming, A.~Krol, and Y.~H. Kao: Phys. Rev. Lett. {\bfseries 66} (1991)
  104.

\bibitem{zhan88}
F.~C. Zhang and T.~M. Rice: Phys. Rev. B {\bfseries 37} (1988) 3759.

\bibitem{peet09}
D.~C. Peets, D.~G. Hawthorn, K.~M. Shen, Y.-J. Kim, D.~S. Ellis, H.~Zhang,
  S.~Komiya, Y.~Ando, G.~A. Sawatzky, R.~Liang, D.~A. Bonn, and W.~N. Hardy:
  Phys. Rev. Lett. {\bfseries 103} (2009) 087402.

\bibitem{chen92}
C.~T. Chen, L.~H. Tjeng, J.~Kwo, H.~L. Kao, P.~Rudolf, F.~Sette, and R.~M.
  Fleming: Phys. Rev. Lett. {\bfseries 68} (1992) 2543.

\bibitem{uchi91}
S.~Uchida, T.~Ido, H.~Takagi, T.~Arima, and Y.~Tokura: Phys. Rev. B {\bfseries
  43} (1991) 7942.

\bibitem{saku11}
Y.~Sakurai, M.~Itou, B.~Barbiellini, P.~E. Mijnarends, R.~S. Markiewicz,
  S.~Kaprzyk, J.-M. Gillet, S.~Wakimoto, M.~Fujita, S.~Basak, Y.~J. Wang,
  W.~Al-Sawai, H.~Lin, A.~Bansil, and K.~Yamada: Science {\bfseries 332} (2011)
  698.

\bibitem{armi10}
N.~P. Armitage, P.~Fournier, and R.~L. Greene: Rev. Mod. Phys. {\bfseries 82}
  (2010) 2421.

\bibitem{yama03}
K.~Yamada, K.~Kurahashi, T.~Uefuji, M.~Fujita, S.~Park, S.-H. Lee, and
  Y.~Endoh: Phys. Rev. Lett. {\bfseries 90} (2003) 137004.

\bibitem{fuji03}
M.~Fujita, S.~Kuroshima, M.~Matsuda, and K.~Yamada: Physica C {\bfseries
  392--396} (2003) 130.

\bibitem{wils06b}
S.~D. Wilson, S.~Li, P.~Dai, W.~Bao, J.-H. Chung, H.~J. Kang, S.-H. Lee,
  S.~Komiya, Y.~Ando, and Q.~Si: Phys. Rev. B {\bfseries 74} (2006) 144514.

\bibitem{moto07}
E.~M. Motoyama, G.~Yu, I.~M. Vishik, O.~P. Vajk, P.~K. Mang, and M.~Greven:
  Nature {\bfseries 445} (2007) 186.

\bibitem{fuji08}
M.~Fujita, M.~Matsuda, S.-H. Lee, M.~Nakagawa, and K.~Yamada: Phys. Rev. Lett.
  {\bfseries 101} (2008) 107003.

\bibitem{mang04a}
P.~K. Mang, O.~P. Vajk, A.~Arvanitaki, J.~W. Lynn, and M.~Greven: Phys. Rev.
  Lett. {\bfseries 93} (2004) 027002.

\bibitem{newm00}
M.~E.~J. Newman and R.~M. Ziff: Phys. Rev. Lett. {\bfseries 85} (2000) 4104.

\bibitem{krug07}
F.~Kruger, S.~D. Wilson, L.~Shan, S.~Li, Y.~Huang, H.-H. Wen, S.-C. Zhang,
  P.~Dai, and J.~Zaanen: Phys. Rev. B {\bfseries 76} (2007) 094506.

\bibitem{tran89b}
J.~M. Tranquada, S.~M. Heald, A.~R. Moodenbaugh, G.~Liang, and M.~Croft: Nature
  {\bfseries 337} (1989) 720.

\bibitem{yama98a}
K.~Yamada, C.~H. Lee, K.~Kurahashi, J.~Wada, S.~Wakimoto, S.~Ueki, Y.~Kimura,
  Y.~Endoh, S.~Hosoya, G.~Shirane, R.~J. Birgeneau, M.~Greven, M.~A. Kastner,
  and Y.~J. Kim: Phys. Rev. B {\bfseries 57} (1998) 6165.

\bibitem{tana05}
Y.~Tanabe, T.~Adachi, T.~Noji, and Y.~Koike: J. Phys. Soc. Jpn. {\bfseries 74}
  (2005) 2893.

\bibitem{wils06}
S.~D. Wilson, P.~Dai, S.~Li, S.~Chi, H.~J. Kang, and J.~W. Lynn: Nature
  {\bfseries 442} (2006) 59.

\bibitem{yu10b}
G.~Yu, Y.~Li, E.~M. Motoyama, K.~Hradil, R.~A. Mole, and M.~Greven: Phys. Rev.
  B {\bfseries 82} (2010) 172505.

\bibitem{zhao07b}
J.~Zhao, P.~Dai, S.~Li, P.~G. Freeman, Y.~Onose, and Y.~Tokura: Phys. Rev.
  Lett. {\bfseries 99} (2007) 017001.

\bibitem{zhao11}
J.~Zhao, F.~C. Niestemski, S.~Kunwar, S.~Li, P.~Steffens, A.~Hiess, H.~J. Kang,
  S.~D. Wilson, Z.~Wang, P.~Dai, and V.~Madhavan: Nat. Phys. {\bfseries 7}
  (2011) 719.

\bibitem{wils06c}
S.~D. Wilson, S.~Li, H.~Woo, P.~Dai, H.~A. Mook, C.~D. Frost, S.~Komiya, and
  Y.~Ando: Phys. Rev. Lett. {\bfseries 96} (2006) 157001.

\bibitem{bour97c}
P.~Bourges, H.~Casalta, A.~S. Ivanov, and D.~Petitgrand: Phys. Rev. Lett.
  {\bfseries 79} (1997) 4906.

\bibitem{fuji06b}
M.~Fujita, M.~Matsuda, B.~F{\aa}k, C.~D. Frost, and K.~Yamada: J. Phys. Soc.
  Jpn. {\bfseries 75} (2006) 093704.

\bibitem{hayd00}
S.~M. Hayden, R.~Doubble, G.~Aeppli, T.~G. Perring, and E.~Fawcett: Phys. Rev.
  Lett. {\bfseries 84} (2000) 999.

\bibitem{tomi87}
S.~Tomiyoshi, Y.~Yamaguchi, M.~Ohashi, E.~R. Cowley, and G.~Shirane: Phys. Rev.
  B {\bfseries 36} (1987) 2181.

\bibitem{yana01}
Y.~Yanase and K.~Yamada: J. Phys. Soc. Jpn. {\bfseries 70} (2001) 1659.

\bibitem{bang09}
Y.~Bang: J. Phys. Conf. Ser. {\bfseries 150} (2009) 052013.

\bibitem{varm97}
C.~M. Varma: Phys. Rev. B {\bfseries 55} (1997) 14554.

\bibitem{varm06}
C.~M. Varma: Phys. Rev. B {\bfseries 73} (2006) 155113.

\bibitem{fauq06}
B.~Fauqu\'e, Y.~Sidis, V.~Hinkov, S.~Pailh\`es, C.~T. Lin, X.~Chaud, and
  P.~Bourges: Phys. Rev. Lett. {\bfseries 96} (2006) 197001.

\bibitem{mook08}
H.~A. Mook, Y.~Sidis, B.~Fauqu\'e, V.~Bal\'edent, and P.~Bourges: Phys. Rev. B
  {\bfseries 78} (2008) 020506.

\bibitem{bale11}
V.~Bal\'edent, D.~Haug, Y.~Sidis, V.~Hinkov, C.~T. Lin, and P.~Bourges: Phys.
  Rev. B {\bfseries 83} (2011) 104504.

\bibitem{bale10}
V.~Bal\'edent, B.~Fauqu\'e, Y.~Sidis, N.~B. Christensen, S.~Pailh\`es,
  K.~Conder, E.~Pomjakushina, J.~Mesot, and P.~Bourges: Phys. Rev. Lett.
  {\bfseries 105} (2010) 027004.

\bibitem{li08}
Y.~Li, V.~Baledent, N.~Barisic, Y.~Cho, B.~Fauque, Y.~Sidis, G.~Yu, X.~Zhao,
  P.~Bourges, and M.~Greven: Nature {\bfseries 455} (2008) 372.

\bibitem{li10b}
Y.~Li, V.~Baledent, G.~Yu, N.~Barisic, K.~Hradil, R.~A. Mole, Y.~Sidis,
  P.~Steffens, X.~Zhao, P.~Bourges, and M.~Greven: Nature {\bfseries 468}
  (2010) 283.

\bibitem{yu10}
G.~Yu, Y.~Li, E.~M. Motoyama, X.~Zhao, N.~Bari\v{s}i\'{c}, Y.~Cho, P.~Bourges,
  K.~Hradil, R.~A. Mole, and M.~Greven: Phys. Rev. B {\bfseries 81} (2010)
  064518.

\bibitem{head11}
N.~S. Headings, S.~M. Hayden, J.~Kulda, N.~H. Babu, and D.~A. Cardwell: Phys.
  Rev. B {\bfseries 84} (2011) 104513.

\bibitem{enok10}
M.~Enoki, M.~Fujita, S.~Iikubo, and K.~Yamada: Physica C {\bfseries 470} (2010)
  S37.

\bibitem{muku08}
H.~Mukuda, Y.~Yamaguchi, S.~Shimizu, Y.~Kitaoka, P.~Shirage, and A.~Iyo: J.
  Phys. Soc. Jpn. {\bfseries 77} (2008) 124706.

\bibitem{webe10}
C.~Weber, K.~Haule, and G.~Kotliar: Nat. Phys. {\bfseries 6} (2010) 574.

\end{thebibliography}

\end{document}